\documentclass[fleqn,usenatbib]{mnras}

\usepackage{newtxtext,newtxmath}

\usepackage[T1]{fontenc}

\DeclareRobustCommand{\VAN}[3]{#2}
\def\peng#1{{\bf \color{blue} PENG: #1}} 
\let\VANthebibliography\thebibliography
\def\thebibliography{\DeclareRobustCommand{\VAN}[3]{##3}\VANthebibliography}

\usepackage{graphicx}	
\usepackage{amsmath}	

\title[]{Turbulent Heating in a Stratified Medium
}

\author[Wang et al.]{
Chaoran Wang,$^{1,3}$\thanks{E-mail: cwn@ucsb.edu}
S. Peng Oh,$^{1}$
M. Ruszkowski$^{2,3}$
\\
${}^{1}$ Department of Physics, University of California, Santa Barbara, CA 93106, USA. \\
${}^{2}$ Max-Planck-Institut für Astrophysik, Karl-Schwarzschild-Stra{\ss}e 1, D-85740 Garching bei München, Germany\\
${}^{3}$ Department of Astronomy, University of Michigan, Ann Arbor, MI 48109, USA
}

\date{Accepted XXX. Received YYY; in original form ZZZ}

\pubyear{2022}

\begin{document}
\label{firstpage}
\pagerange{\pageref{firstpage}--\pageref{lastpage}}
\maketitle

\def\lsim{\;\rlap{\lower 2.5pt
   \hbox{$\sim$}}\raise 1.5pt\hbox{$<$}\;}

\begin{abstract}
There is considerable evidence for widespread subsonic turbulence in galaxy clusters, most notably from {\it Hitomi}. Turbulence is often invoked to offset radiative losses in cluster cores, both by direct dissipation and by enabling turbulent heat diffusion. However, in a stratified medium, buoyancy forces oppose radial motions, making turbulence anisotropic. This can be quantified via the Froude number ${\rm Fr}$, which decreases inward in clusters as stratification increases. We exploit analogies with MHD turbulence to show that wave-turbulence interactions increase cascade times and reduces dissipation rates $\epsilon \propto {\rm Fr}$. Equivalently, for a given energy injection/dissipation rate $\epsilon$, turbulent velocities $u$ must be higher compared to Kolmogorov scalings. High resolution hydrodynamic simulations show excellent agreement with the $\epsilon \propto {\rm Fr}$ scaling, which sets in for ${\rm Fr} \lsim 0.1$. We also compare previously predicted scalings for the turbulent diffusion coefficient $D \propto {\rm Fr}^2$ and find excellent agreement, for ${\rm Fr} \lsim 1$. However, we find a different normalization, corresponding to stronger diffusive suppression by more than an order of magnitude. Our results imply that turbulent diffusion is more heavily suppressed by stratification, over a much wider radial range, than turbulent dissipation. Thus, the latter potentially dominates. 
Furthermore, this shift implies significantly higher turbulent velocities required to offset cooling, compared to previous models. These results are potentially relevant to turbulent metal diffusion (which is likewise suppressed), and to planetary atmospheres. 

\end{abstract}


\newcommand{\noga}{NoG-1/2}
\newcommand{\nogb}{NoG-5/6}
\newcommand{\nogc}{NoG-4/3}
\newcommand{\per}{PER}
\newcommand{\stra}{STR-eps-high}
\newcommand{\strb}{STR-eps-medium}
\newcommand{\strc}{STR-eps-low}
\newcommand{\strg}{STR-10G}
\newcommand{\bv}{Brunt-Väisälä}
\def\peng#1{{\bf \color{blue} PENG: #1}} 
\def\chaoran#1{{\bf \color{red}  Chaoran: #1}} 
\def\mateusz{{\bf \color{red} MATEUSZ: }} 

\section{Introduction}

The intracluster medium (ICM) is the dominant baryonic component filling the vast volume of galaxy clusters. In the form of hot ionized plasmas ($T\sim$keV), the ICM has been extensively 
observed in X-ray. These observations of the ICM have revealed the prevalence of  turbulence, using indirect methods based on surface brightness fluctuations 
\citep{2013A&A...559A..78G, 2014Natur.515...85Z}, resonance scattering \citep{2017MNRAS.472.1659O}, and Sunyaev-Zeldovich (SZ) fluctuations \citep{2012ApJ...758...74B}. However, not until the launch of the {\it Hitomi} telescope were direct measurements of the ICM turbulence viable, as early X-ray spectroscopy was limited by spatial and energy resolution \citep[e.g.,][]{Sanders2013}. The first direct observation of turbulence obtained by the Hitomi telescope measured the Doppler line broadening of the Fe XXV and Fe XXVI emission lines in the ICM of the Perseus cluster \citep{Hitomi2016}. Recently, \citet{Li2020} directly probed the turbulence in the cold ICM by measuring the velocity structure functions of the cold ICM filaments 
in the very central regions of three nearby clusters
using optical spectroscopic data. The amplitude of velocity fluctuations of cold filaments is comparable to that of the hot medium, indicating the hot and cold phases are dynamically coupled, agreeing with the numerical simulations of the multiphase ICM \citep{Wang21}.  
Overall, these measurements all find that the ICM turbulence is subsonic and the turbulent energy density is very small compared to the thermal pressure of the ICM. 

There is a consensus that feedback from the active galactic nuclei (AGN) associated with the central supermassive black hole can balance the observed radiative cooling, hence maintaining the global thermal equilibrium of the hot gaseous halos (see \citealt{2007ARA&A..45..117M, 2012NJPh...14e5023M} and \citealt{2012ARA&A..50..455F} for reviews). However, how the  AGN energy is coupled with the gaseous halo remains an open question. Volume-filling turbulence could play a role, either by dissipation of gas motions into heat, or by facilitating heat transport from the high entropy cluster outskirts (which serves as a heat bath due to its long cooling time) to the lower entropy cluster core, where most cooling takes place. 
For turbulent dissipation to play a role in  thermal equilibrium, the heating rate due to turbulence must be comparable to the observed radiative cooling rate. By assuming a one-to-one conversion between density fluctuations observed in residual X-ray brightness map to the velocity fluctuations caused by turbulence,
\citet{2014Natur.515...85Z} derive the velocity power spectra  of the hot ICM in the Perseus and Virgo clusters. The resultant velocity power spectra are broadly consistent with the \citet{kolmogorov41} prediction, i.e., $v_l\propto l^{1/3}$. 
Based on the derived velocity power spectra, \citet{2014Natur.515...85Z} estimate the turbulent dissipation rate $v_l^3/l$, and find that it can balance the radiative cooling rate as a function of the distance from cluster centers. Therefore, their results suggest that turbulent dissipation can be the dominant mechanism for energy transfer from AGN outflows to the hot ambient ICM. 

Alternatively, turbulence can facilitate energy transport from large radii. Mixing of gas due to turbulence leads to net radial entropy inflow in the ICM, since the ICM has a universal positive entropy gradient \citep[e.g.,][]{2018ApJ...862...39B}. This process, known as turbulent diffusion, results in additional heating on top of turbulent dissipation. 
Using analytical models, \cite{dennis05} (hereafter DC05) find turbulence with velocity dispersion in the range of 100$\sim$300 km s$^{-1}$ can balance radiative cooling. They find that both turbulent diffusion and dissipation are energetically important. Similar results are found by \cite{Fujita20}, where the unstable balance between turbulent heating and radiative cooling is dynamically sustained by the modulation from AGN feedback.

However, subsequent studies have cast doubt on the significance of turbulence in heating the ICM. For example, by statistically analyzing a large set of hydrodynamical cluster simulations,  \citet{2019ApJ...874...42V} find that the turbulent velocity is too small to make significant contribution to the thermal energy budget of the cluster cores. Similar results were reported by other teams simulating self-regulated AGN feedback in the ICM \citep{Li2017, Yang2016}. Moreover, \citet{2019MNRAS.484.4881M} find that for the dissipation of subsonic turbulence to balance radiative losses, the turbulent mixing time must be shorter than radiative cooling time. This would imply that no thermal instability could operate in the cool core, contrary to what is observed. That density fluctuations are proportional to the turbulent velocity fluctuations, as adopted in \cite{2014Natur.515...85Z} is also questioned. As shown by the simulations in \citet{2020MNRAS.493.5838M}, the gravitational stratification can affect the amplitude of density fluctuations. Generally, strong stratification leads to larger density fluctuations for a given turbulent velocity. Thus, density fluctuations can overestimate the velocity fluctuations and hence the turbulent dissipation rate in the strongly stratified ICM and in particular in the central region, 
where stratification is the strongest.
Furthermore, density fluctuations can arise due to factors other than turbulence, such as, e.g., contact discontinuities at the boundary between AGN-inflated bubbles and the ambient ICM.

The properties of turbulence are altered in stratified medium. Subsonic gas motions in stratified ICM sustain internal gravity waves \citep{Ruzkowski10}. Therefore, turbulence in stratified ICM can be understood as wave turbulence where nonlinear interactions of internal waves lead to a turbulent cascade. Such a cascade via wave-wave interactions is quite different from the usual Kolmogorov picture. For instance, the cascade time is no longer given by the eddy turnover time $\tau_{\rm eddy} \sim l/u$, where $u$ and $l$ are the characteristic velocity and scale length of turbulence, respectively; thus, the volumetric turbulent dissipation rate is {\it not} $u^{3}/l$. The aim of this paper is to study how this alters turbulent dissipation and turbulent heat diffusion rates, and how this affects the contribution of turbulence to thermodynamic energy balance in galaxy clusters. 

The influence of stratification on the turbulent heating rate has not been investigated in the astrophysical literature. In the context of the Earth's atmosphere, it was studied numerically by \citet{pouquet18}. They directly measured the actual turbulent dissipation rate $\epsilon$ and compared it with the Kolmogorov expectation $u^{3}/L$. They found that $\beta \equiv \epsilon/(u^{3}/L) \propto$ Fr, where Fr$\equiv u/(L N)$ is the Froude number, and $N$ is the \bv\  frequency. The Froude number measures the importance of stratification; as stratification increases, the Froude number falls. Thus, \citet{pouquet18} found that stratification reduces the efficacy of turbulent dissipation. However, they did not provide a physical explanation for this scaling. In this paper, we derive this scaling analytically, which to our knowledge has not been done before, and show numerically that it also holds for galaxy clusters. 

Additionally, gravitational stratification suppresses turbulent diffusion. Turbulent diffusion parallel to gravity is suppressed, because turbulent velocities are reduced in this direction. This {\it has} been accounted for in the astrophysical literature. DC05 and \citet{Fujita20} utilize the analytical model for stratified turbulent diffusion of \citet{weinstock81}, which was originally derived in the context of the Earth's atmosphere. However, these models have not been tested numerically, at least in the context of galaxy clusters.
Furthermore, the models of DC05 and \citet{Fujita20} make certain assumptions which may not necessarily hold. The characteristic scale of turbulence is set to be proportional to the distance from the ICM center ($r$): $l \propto r$. 
We shall argue that this need not be the case: the characteristic scale of turbulence could be much larger, without scaling with $r$; this implies a much lower Fr and hence stronger stratification. Our numerical simulations find that supression of heat diffusion can be stronger by more than an order of magnitude than previously estimated, both due to dimensionless numerical coefficients (which must be calibrated to simulations) and differing assumptions about the turbulent driving scale. 

\citet{weinstock81} and \citet{pouquet18} provide theoretical models describing how the stratification affects the rate of heating due to turbulence. However, these models are derived based on numerical simulations with conditions consistent with the Earth's atmosphere, which are not appropriate for the galaxy cluster environment. These conditions include but are not limited to: plane-parallel geometry, namely, vertical stratification; an isothermal equation of state. In the intracluster medium, the curvature of the central regions cannot be ignored; the  gas is quasi-adiabatic rather than isothermal. All these differences suggest the necessity of coming up with new theoretical models suitable for understanding the thermodynamic influence of turbulence in galaxy clusters.

The outline of this paper is as follows. In \S\ref{sec:analytics}, we describe analytic models for the impact of stratification on turbulent dissipation and heat transport. In \S\ref{sec:methods}, we describe our methodology for performing high resolution 3D hydrodynamic simulations of driven turbulence in galaxy clusters. In \S\ref{sec:results}, we describe our results and confront analytic models with simulations results. We discuss and conclude in \S\ref{sec:conclude}.  

\begin{table*}
	\centering
	\caption{List of Simulations}
	\label{tab:listOfSims}
	\begin{tabular}{lcccccccr} 
		\hline
		name  & $L_{\rm box}$(kpc)  & $\Delta x$(kpc)            & $\epsilon_{\rm inj}{(\rm erg/g/s})$  &  {\rm Fr} & Entropy profile & Gravity                & $l_{\rm drive}$(kpc)        & $u_{\rm rms}$(km/s) \\
		(1)&(2)&(3)&(4)&(5)&(6)&(7)&(8)&(9)\\
		\hline
		\noga & 500                 & 1.526 & $8\times10^{-6}$   & $\infty$              & $\propto r^{1/2}$     & no gravity             & $20$                    &  $135$          \\
		\nogb & 500                 & 1.526 & $8\times10^{-6}$   & $\infty$              & $\propto r^{5/6}$     & no gravity             & $20$                    &  $126$              \\
		\nogc & 500                 & 1.526 & $8\times10^{-6}$   & $\infty$              & $\propto r^{4/3}$     & no gravity             & $20$                    &  $107$      \\
		\per  & 500                 & 1.526 & $8\times10^{-6}$   & $0.4 \sim0.9$         & Perseus                     & NFW+stellar            & $20$                    &  $109$      \\
		\stra & 125                 & 0.39  & $6.4\times10^{-7}$ & $0.04\sim0.1$         & Universal                   & NFW+stellar            & $50$                    &  $70$               \\
		\strb & 125                 & 0.39  & $3.2\times10^{-7}$ & $0.03\sim0.08$        & Universal                   & NFW+stellar            & $50$                    &  $55$       \\
		\strc & 125                 & 0.39  & $1.2\times10^{-7}$ & $0.02\sim0.06$        & Universal                   & NFW+stellar            & $50$                    &  $43$       \\
		\strg & 125                 & 0.39  & $3.2\times10^{-7}$ & $0.01\sim0.03$        & 10$\times$Universal         & 10$\times$(NFW+stellar)& $50$                    &  $70$               \\
		\hline
	\end{tabular}
	\\ \ \\
 \raggedright{{\bf Note} (1) name of the runs, (2) size of the simulation domain, (3) spatial resolution, (4) normalization of the spectral forcing scheme, which approximately measures the energy injected per unit mass per mode by the spectral forcing (see Section~\ref{sec:forcingScheme} for details), (5) range of azimuthally-averaged {\rm Fr}, (6) choice of initial entropy profiles. For NoG- runs, entropy profiles are initially set to be power laws; for the \per\ run, the entropy profile of the Perseus cluster is adopted; for the STR- runs, the entropy profiles are set up based on an universal entropy profile. See section~\ref{sec:initialCondition} for details. (7) choice of gravitational potential. For NoG- runs, no gravity is included; for all other runs, gravity is contributed by an NFW profile of dark matter mass and the mass of the stars. See section~\ref{sec:initialCondition} for details. (8) driving scale of the turbulence, (9) the velocity dispersion of turbulence.} 
\end{table*}
\section{Stratified Turbulence: Analytic Expectations} 
\label{sec:analytics} 

\subsection{Turbulent Dissipation} \label{sec:stratTurb}

Turbulence in a stratified medium is anisotropic, and has close parallels with MHD turbulence. In both cases, the system has a preferred direction singled out by gravity and magnetic fields respectively, and supports linear waves which can interact with and modify turbulence. The wave frequency $\omega=N$ (where $N$ is the Brunt-Väisälä frequency) competes against the non-linear decorrelation rate $\tau_{\rm NL}^{-1} \sim u/L$ set by the non-linear advection term $u \cdot \nabla u$ in the Euler equation; the ratio of these frequencies ${\rm Fr} = u/(NL) \approx (\omega \tau_{\rm NL})^{-1}$ is known as the Froude number. 

Turbulence is characterized by a constant energy flux $\epsilon \approx \langle u^2 \rangle/\tau_{\rm cas}$ across scales, where $\tau_{\rm cas}$ is the cascade time. In steady state, this equals the dissipation rate. Our goal here is to estimate the impact of stratification on $\tau_{\rm cas}$, and in particular to find the scaling relation between $\tau_{\rm cas}$ and the Froude number Fr. This requires understanding the interaction between waves and turbulence. The discussion here is approximate and qualitative; for excellent reviews and more rigorous reviews of wave turbulence, see \citet{zakharov92,nazarenko11-book,nazarenko11}. 

From Kolmogorov turbulence, we are used to thinking of the cascade time as equal to the non-linear decorrelation (eddy turnover) time, $\tau_{\rm cas} \sim \tau_{\rm NL} \sim L/u$. This is intuitively reasonable; for systems with only turbulent eddies, there is no other timescale in the problem. However, in a system which supports both waves and turbulence, there is another timescale associated with the wave frequency $\omega$. How does this affect $\tau_{\rm cas}$? Also, the presence of a mean field such as a B-field, or gravity, introduces anisotropy $k_{\perp}/k_{\parallel} \neq 1$. How does anisotropy affect $\tau_{\rm cas}$? Equivalently, the energy power spectrum $E(k)$, defined such that $\langle u^2 \rangle \sim \int dk E(k) \sim k E(k)$, must change in a system that supports waves. In Kolmogorov theory, $E(k) \sim \epsilon^{2/3} k^{-5/3}$ is usually derived via dimensional analysis from $\epsilon,k$. However, once waves are present, a third parameter which characterizes them (such as the Alfven speed $v_A$ in MHD turbulence, or the Brunt-Väisälä frequency $N$ in stratified turbulence) now appears. Due to this additional parameter, the system becomes degenerate, and it is no longer possible to uniquely determine $E(k)$ from dimensional analysis. Note that $\tau_{\rm cas}$ and $E(k)$ are related, since 
\begin{equation} 
\epsilon \sim \frac{\langle u^2 \rangle}{\tau_{\rm cas}} \sim \frac{k E(k)}{\tau_{\rm cas}}.  
\label{eqn:tau_cas}  
\end{equation}

Getting out of this impasse requires an additional closure relation. The appropriate closure depends on $\omega \tau_{\rm NL}$, where $\omega \tau_{\rm NL} \sim M_A^{-1}$ for Alfvenic turbulence (using $\omega \sim v_A k$, and $\tau_{\rm NL} \sim (u k)^{-1}$), and $\omega \tau_{\rm NL} \sim {\rm Fr}^{-1}$ for stratified turbulence. In the $\omega \tau_{\rm NL} \ll 1$ regime, turbulence is strong, and waves are only a small perturbation; Kolmogorov turbulence is appropriate. Conversely, in the $\omega \tau_{\rm NL} \gg 1$ regime, turbulence and non-linearity is weak. This weak turbulence regime is the most relevant for us; it is equivalent to the case when stratification is strong, i.e., when ${\rm Fr} \sim (\omega \tau_{\rm NL})^{-1} \ll 1$. In this case, non-linearity can be treated perturbatively in the small parameter $(\omega \tau_{\rm NL})^{-1}$, which allows one to calculate $E(k)$ and $\tau_{\rm cas}$ \citep{zakharov92}. An important caveat is that $\omega \tau_{\rm NL}$ is a function of scale. Non-linearity generally increases towards small scales, as $\tau_{\rm NL}$ falls. The celebrated critical balance hypothesis \citep{goldreich95,nazarenko11} states that the system will tend toward a state where $\omega \tau_{\rm NL} \sim 1$, i.e., there is a scale-by-scale balance between linear propagation times and non-linear interaction times over a wide range of scales. For the case of MHD turbulence (where Goldreich and Sridhar first introduced it), critical balance has considerable evidence both in numerical simulations \citep{cho00,maron01}, as well as solar wind data \citep{horbury08,podesta09,wicks10,chen11}. The critical balance hypothesis replaces two timescales with a single timescale $\tau_{\rm NL} \sim \omega^{-1}$, and therefore also fixes $E(k),\tau_{\rm cas}$. The lifting of degeneracy also allows one to calculate anisotropy $k_{\perp}/k_{\parallel}$ as a function of scale.

{\bf Weak turbulence (strong stratification).} First, let us consider velocity anisotropy. As stratification increases (Fr$\rightarrow 0$), the restoring forces in the vertical direction become stronger, and gas motions are increasingly confined to 2D planes perpendicular to the direction of gravity, such that $k_{\perp} \ll k \approx k_{\parallel}$, i.e., fluid motions are strongly anisotropic. From incompressibility (assuming subsonic turbulence) $\nabla \cdot u =0$, we obtain $u_{\perp} \sim (k_{\parallel}/k_{\perp}) u_{\parallel} \gg u_{\parallel}$. Thus, most of the kinetic energy is in the perpendicular direction (i.e., in horizontal motions), and the non-linear cascade proceeds primarily in the perpendicular direction. We can rewrite equation \ref{eqn:tau_cas} as: 
\begin{equation} 
\epsilon \sim \frac{\langle u_{\perp}^2 \rangle}{\tau_{\rm cas}} \sim \frac{k_{\perp} E(k_{\perp})}{\tau_{\rm cas}}. 
\label{eqn:tau_cas_perp}  
\end{equation} 

Furthermore, in the strongly stratified, weak turbulence limit ${\rm Fr} \ll 1$, the power spectrum is: 
\begin{equation} 
E(k_{\perp}) \sim (N \epsilon)^{1/2} k_{\perp}^{-2}. 
\label{eqn:Ek-weak-turb} 
\end{equation} 
The $E(k) \propto k^{-2}$ scaling was first reported from a empirical fit to oceanographic measurements in a famous paper \citep{garrett75}, and has been rigorously derived via a Hamiltonian/kinetic equation approach (e.g., \citet{pelinovsky77,caillol00, lvov01}); similar spectra have also been derived for weak turbulence in the MHD context \citep{galtier00}, and verified numerically \citep{perez08}. If we insert equation \ref{eqn:Ek-weak-turb} into equation \ref{eqn:tau_cas_perp}, we obtain: 
\begin{equation} 
\tau_{\rm cas} \sim \frac{1}{(k_{\perp} u_{\perp})^{2}} N \sim \tau_{\rm NL} \left(\frac{N}{ku} \right) \sim \frac{\tau_{\rm NL}}{{\rm Fr}} \ \ ({\rm Fr} \ll 1)
\label{eqn:tau_cas_Fr} 
\end{equation} 

Equation \ref{eqn:tau_cas_Fr} is the main result of this section. Recall that cascade times $\tau_{\rm cas}(k_{\perp})$ are a function of scale, decreasing as one goes to smaller scales. Furthermore, as the cascade proceeds to smaller scales and $\omega \tau_{\rm NL}$ falls, it will enter the critical balance regime, which have different power spectra and cascade times $\tau_{\rm cas}$ (see below). However, since measured velocities are dominated by the outer scale, we want to know the cascade time at the outer scale as well (i.e., the maximum value of $\tau_{\rm cas}$), in order to accurately determine the dissipation rate:
\begin{equation} \label{eq:dissScaling}
\epsilon \sim \frac{\langle u^2 \rangle}{\tau_{\rm cas}} \sim \frac{\langle u^2 \rangle}{\tau_{\rm NL}} 
\ {\rm Fr} \sim \epsilon_{K} \ {\rm Fr} \sim \frac{u^4}{N L^{2}} 
\end{equation} 
where $\epsilon_{\rm K}$ is the usual Kolmogorov dissipation rate. Thus, for a measured velocity dispersion $\langle u^2 \rangle$, the longer cascade time implies a {\it decreased} energy dissipation rate, by a factor Fr. Using Kolmogorov scalings overestimates turbulent dissipation by ${\rm Fr}^{-1}$, which can be up to an order of magnitude in clusters. Furthermore, since now $\epsilon \propto L^{-2}$ instead of $\epsilon \propto L^{-1}$, one is very sensitive to the assumed driving scale $L$, which is not directly measured. Similarly, since now $\epsilon\propto u^{4}$ rather than $\epsilon\propto u^{3}$, dissipation estimates are more sensitive to $u$ in stratified media.
Conversely, for a fixed energy injection rate $\epsilon$, the longer cascade time implies an {\it increased} velocity dispersion: 
\begin{equation}\label{eq:vFr}
v \sim \left( {\epsilon N}{L^{2}}  \right)^{1/4} \sim \left( \frac{\epsilon L}{\rm Fr} \right)^{1/3}.  
\end{equation} 
by a factor ${\rm Fr}^{-1/3}$, compared to canonical Kolmogorov values $\epsilon \sim v^{3}/L$, $v \sim (\epsilon L)^{1/3}$. 

Since we did not derive the power spectrum (equation \ref{eqn:Ek-weak-turb}), it is worth understanding equation \ref{eqn:tau_cas_Fr} from another angle, via random walk arguments originally developed in the MHD context \citep{nazarenko11}. In the weak turbulence limit, waves are the fundamental modes of the system, and non-linearities occur when wave packets collide and subsequently distort. Since they interact on the short wave crossing time $\tau_{\rm BV} = N^{-1}$ rather than the eddy turnover time $\tau_{\rm eddy} \sim L_{\perp}/v_{\perp}$, with $\tau_{\rm BV} \ll \tau_{\rm eddy}$, the non-linear interaction can be treated perturbatively. Each interaction results in the small velocity change $\delta u_{\perp} \sim u_{\perp} (\tau_{\rm BV}/\tau_{\rm eddy}) \ll u$. These small, uncorrelated velocity changes will sum like a random walk. A given number of interactions over time $\tau$, $n \sim \tau/\tau_{\rm BV}$, where the frequency of wave packet collisions is $\sim \tau_{\rm BV}^{-1}$, 
will produce a net velocity perturbation $\Delta u_{\perp} \sim n^{1/2} \delta u_{\perp} \sim (\tau/\tau_{\rm BV})^{1/2} (\tau_{\rm BV}/\tau_{\rm eddy}) u$. The cascade time $\tau_{\rm cas}$ can be defined as the timescale on which $\Delta u \sim u$, i.e. the perturbation grows non-linear and cascades to smaller scales. Solving the expression $\Delta u \sim u$ for $\tau_{\rm cas}$, we obtain: 
\begin{equation} 
\tau_{\rm cas} \sim \left( \frac{\tau_{\rm eddy}}{\tau_{\rm BV}} \right)^{2} \tau_{\rm BV} \sim \frac{L N}{u} \tau_{\rm eddy} \sim \frac{\tau_{\rm eddy}}{{\rm Fr}}, 
\end{equation} 
which agrees with equation \ref{eqn:tau_cas_Fr}. The power spectrum $E(k_{\perp})$ (equation \ref{eqn:Ek-weak-turb}) can then be derived from the above expression $\tau_{\rm cas} \sim \tau_{\rm NL}/$Fr and equation \ref{eqn:tau_cas_perp} . 

Finally, the power spectrum $E(k_{\perp})$ can be derived from dimensional analysis if an additional constraint, the fact that stratified turbulence is a `3 wave process' (two waves collide to produce a third wave), with the energy $E$ satisfying $\epsilon \sim \dot{E} \sim E^2$ (similar to a binary chemical reaction; in general, for an $n$ wave process, $\epsilon \sim \dot{E} \sim E^{n-1}$), is imposed \citep{nazarenko11-book}; this reasoning can be justified from the form of the wave kinetic equation. This imposes the additional constraint $E(k) \propto E \propto \epsilon^{1/2}$, which lifts the degeneracy which arose from introducing an additional parameter, the Brunt-Vaisala frequency $N$. Performing dimensional analysis on $E(k) \sim N^{a} \epsilon^{1/2} k^{-b}$, where $[E(k)] = L^3 T^{-2}$, $[\epsilon]=L^2 T^{-3}$, $[N] = T^{-1}$, $[k]=L^{-1}$, gives $a=1/2$, $b=-2$, recovering equation \ref{eqn:Ek-weak-turb}. While this n-wave reasoning (first developed by \citealt{kraichnan65}; most waves are n=3 ($2\rightarrow 1$) or n=4 ($2\rightarrow2$)) seems crude, it appears to recover the correct power spectra for most wave systems, including Alfven waves, waves in rotating fluids, water gravity waves, Langmuir waves.  

{\bf Critical Balance.} For completeness, it is useful to consider scalings once the system reaches critical balance, $\omega \tau_{\rm NL} \sim 1$. The system remains anisotropic, with the cascade proceeding primarily in the perpendicular (i.e., horizontal) direction. However, since $\omega \sim N \sim \tau_{\rm NL}^{-1}$, there is once again only one timescale in the system, as in Kolmogorov turbulence. Since $\tau_{\rm cas} \sim \tau_{\rm NL} \sim N^{-1}$, then $\epsilon \sim u^{2}/\tau_{\rm cas} \sim u^{2} N$, i.e. $\epsilon, N$ are no longer independent parameters. The choice of key parameter depends on direction. In the perpendicular direction, where the turbulent cascade operates, $\epsilon$ must be the relevant parameter. This is identical to the situation in Kolmogorov turbulence, and once again by dimensional analysis, the spectrum is Kolmogorov: $E(k_{\perp}) \sim \epsilon^{2/3} k_{\perp}^{-5/3}$. In the parallel direction, where wave motions dominate, the relevant dimensional parameter is $N$. By dimensional analysis, we obtain $E(k_{\parallel}) \sim N^2 k_{\parallel}^{-3}$. Finally, with a bit more care in evaluating the dispersion relation for buoyant oscillations, $\omega \sim N k_{\perp}/k_{\parallel}$, we can evaluate anisotropy. Critical balance, $\omega \sim \tau_{\rm NL}^{-1}$, where $\omega \sim N k_{\perp}/k_{\parallel}$ and $\tau_{\rm NL}^{-1} \sim k_{\perp} u_{\perp}$, combined with the Kolmogorov scaling $u_{\perp} \sim \epsilon^{1/3} k_{\perp}^{-1/3}$, gives the relation: 
\begin{equation}
k_{\perp} \sim \frac{\epsilon}{N^{3}} k_{\parallel}^{3}  \sim l_{\rm O}^{2} k_{\parallel}^{3}
\label{eqn:CB-stratified}
\end{equation} 
where $l_{\rm O} \sim (\epsilon/N^{3})^{1/2}$ is the Ozmidov scale. This is analogous to the well-known MHD critical balance condition, $v_{\rm A} k_{\parallel} \sim \epsilon^{1/3} k_{\perp}^{2/3}$, where the turbulent cascade is also primarily perpendicular. However, there are critical differences. In Alfvenic turbulence, restoring forces are in the perpendicular direction, and waves propagate in the parallel direction; we have $k_{\perp} \gg k_{\parallel}$. In stratified turbulence, restoring forces are in the parallel direction, and waves propagate in the perpendicular direction; we have $k_{\parallel} \gg k_{\perp}$, i.e., the roles of $k_{\perp},k_{\parallel}$ are switched. Note that for Alfvenic turbulence, $k_{\parallel}/k_{\perp} \propto k_{\perp}^{-1/3}$, while for stratified turbulence, $k_{\parallel}/k_{\perp} \propto k_{\perp}^{-2/3}$, i.e. in both cases $k_{\parallel}/k_{\perp}$ decreases towards small scales. However, in MHD turbulence, the cascade begins with approximate isotropy ($k_{\parallel} \sim k_{\perp}$ at large scales, when MHD forces are weaker) and evolves towards anisotropy ($k_{\parallel} \ll k_{\perp}$) at small scales, while in stratified turbulence, gas motions are highly anisotropic at large scales ($k_{\parallel} \gg k_{\perp}$), and evolve toward isotropy at small scales, becoming isotropic ($k_{\parallel} \sim k_{\perp}$) at the Ozmidov scale, $k \sim l_{\rm O}^{-1}$.
This makes sense: on small scales, the system looks uniform, and stratification is unimportant. At scales smaller than the Ozmidov scale, turbulence is an isotropic Kolmogorov cascade.

In our simulations, $\epsilon = \langle a \cdot v \rangle$ is the energy injection rate. However, not all of this is deposited as heat; some fraction is deposited as potential energy. In both observational data in the stratosphere \citep{lindborg06} and simulations \cite{pouquet18}, the ratio of potential energy to kinetic energy injection is inferred to be $f_{\rm pot} \sim \epsilon_{\rm P}/\epsilon_{\rm V} \sim 0.3$. 
In general we expect (and our simulations are consistent with) quasi-equipartition values, $f_{\rm pot} \sim \mathcal{O}(1)$, for virialized gas. Thus, the heating rate is $\sim {\rm Fr} (1-f_{\rm pot}) (u^3/L)$.

Cascade times (and hence heating rates) in the weak turbulence limit can also be affected by rotation, as parameterized by the Rossby number. This is beyond the scope of this work. 

\subsection{Turbulent Heat Diffusion}
In a strongly stratified medium, turbulent heat diffusion arises due to small scale vertical motions, where wave motions dominate. The amplitude of vertical oscillations is: 
\begin{equation}
    l_{\parallel} \sim \frac{u}{N} \sim L \left( \frac{u}{N L} \right) \sim ({\rm Fr}) \, L,
\end{equation}
i.e., the anisotropy scales with Froude number, $l_{\parallel}/L \sim {\rm Fr}$. While this is intuitive, the vertical scale $l_{\parallel} \sim {u}/{N}$ can also be formally obtained via a similarity analysis, from the fact that the fluid equations in a strongly stratified medium (${\rm Fr} \ll 1$) are self-similar with respect to the variable $z N/u$ \citep{billant01}, and it has also been demonstrated in numerical simulations \citep{lindborg06}. At the same time, incompressibility gives: 
\begin{equation} 
u_{\parallel} \sim \left( \frac{k_{\perp}}{k_{\parallel}} \right) u_{\perp} \sim \left(\frac{l_{\parallel}}{L} \right) u \sim ({\rm Fr}) \, u. 
\end{equation} 
Putting this together, we obtain: $D_{\parallel} \sim v_{\parallel} l_{\parallel} \sim ({\rm Fr})^{2} u L$, i.e., the diffusion coefficient is suppressed by a factor ${\rm Fr}^{2}$ in strongly stratified settings.  
If we smoothly interpolate between the unstratified and stratified regimes, we can write (\citealt{weinstock81}; DC05): 
\begin{equation}\label{eq:stratDiff}
    D_{\parallel} = c_{0}  \frac{u L}{1+ c_1 {\rm Fr}^{-2}}
\end{equation}
where $c_{0}, c_{1}$ are dimensionless constants of order unity. 

\begin{figure*}
     \centering
    \includegraphics[width=\textwidth]{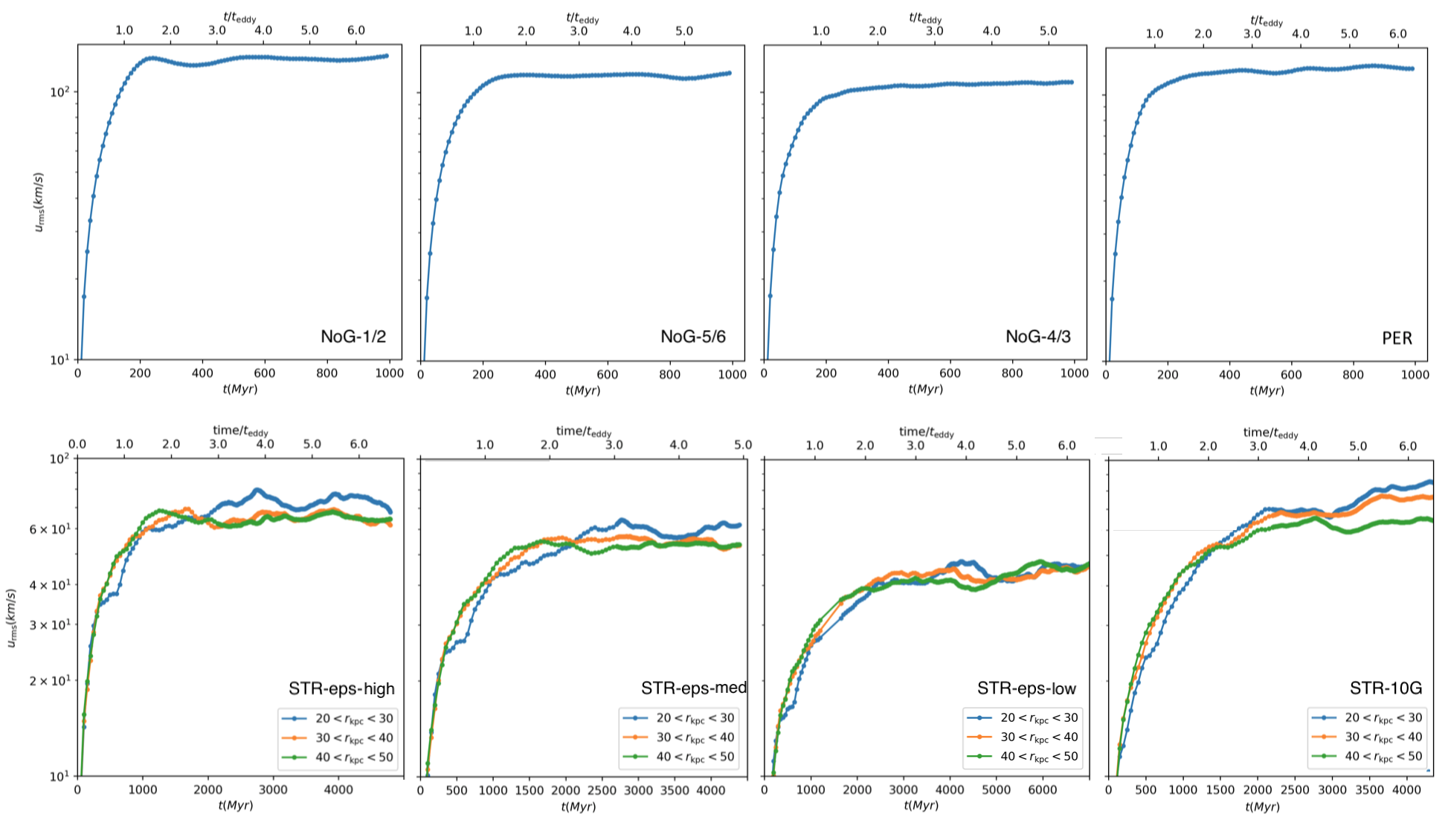}  
       \caption[]{Time evolution of the 3D velocity dispersion, $u_{\rm rms}(t)$. The no-gravity and weakly-stratified cases are in the top row and the strongly-stratified cases are in the bottom. Name of the run is labelled in each panel. For the no-gravity and weakly-stratified cases, we show $u_{\rm rms}(t)$ calculated over the entire volume; and for the strongly-stratified cases, we show $u_{\rm rms}(t)$ in three radial shells, $20<r_{\rm kpc}<30$, $30<r_{\rm kpc}<40$ and $40<r_{\rm kpc}<50$. The upper ticks of each panel show the simulation time scaled by the eddy-turn over time, $t_{\rm eddy} = L/u_{\rm rms} $.}
     \label{fig:sigAll}
\end{figure*}

\section{Methodology}
\label{sec:methods} 
We perform simulations using the FLASH code \citep{2000ApJS..131..273F,dubey2008introduction}. For all runs performed, the simulation domain is resolved by 320$^3$ zones. We adopt the uniform grid mode to avoid inhomogeneity of numerical dissipation. 

We set up the simulation domain with a static gravitational potential, and with gas in hydrostatic equilibrium with a specified density and temperature profile. 
In order to better isolate the effect of turbulent heating, we turn off radiative cooling for all simulations. Thus, all entropy changes can be attributed to the effects of turbulence. 
The system evolution is then governed by the following set of hydrodynamic equations with source terms corresponding to the spectral forcing scheme of driving turbulence, ${\bf a}_{\rm turb}$. 

\begin{equation}
    \frac{\partial \rho}{\partial t} + \nabla\cdot(\rho {\bf v}) = 0;
\end{equation}
\begin{equation}
    \frac{\partial (\rho {\bf v}) }{\partial t} + \nabla\cdot(\rho {\bf v}{\bf v}) = \rho({\bf g} - \nabla p + {\bf a_{\rm turb}});
\end{equation}
\begin{equation}
    \frac{\partial e_{\rm tot} }{\partial t} + \nabla\cdot[(e_{\rm tot}+p ) {\bf v}] = \rho({\bf g}  + {\bf a_{\rm turb}})\cdot {\bf v},
\end{equation}
where $\rho$, $p$, ${\bf v}$ are the gas density, thermal pressure, and velocity, respectively; $e_{\rm tot}$ is the sum of kinetic and internal energy of the gas; and ${\bf g}$ is the gravitational acceleration.
In section~\ref{sec:forcingScheme} we explain how ${\bf a}_{\rm turb}$ is calculated. The initial setup for all performed simulations is described in section~\ref{sec:initialCondition}.

\subsection{Spectral forcing scheme}\label{sec:forcingScheme}
To include the turbulence source terms, we adopt the ``Stir Unit'' with the ``Generate'' implementation in FLASH utilizing a spectral forcing scheme. The numerical method is described in detail in  \citet{2010A&A...512A..81F}. Here we briefly summarize the key points. 

The forcing field is calculated in the spatial Fourier space. Each Fourier mode of each spatial component of the vector field, $a^{(x, y, z)}_{\bf k}(t)$ is evolved independently by an Ornstein-Uhlenbeck random process:
\begin{equation}\label{eq:ou}
    a^{(x, y, z)}_{\bf k}(t+\Delta t) = f a^{(x, y, z)}_{\bf k}(t) + \sigma \sqrt{1-f^2} Z ,
\end{equation}
where $f={\rm exp}(-\Delta t/t_{\rm decay})$ is a damping factor; $\sigma=\sqrt{\epsilon_{\rm inj}/t_{\rm decay}}$ is the desired variance of $a_{\bf k}$; $Z\sim N(0, 1)$ 
is a standard normal random variable, with zero mean and standard deviation of 1; and the initial condition is 
\begin{equation}\label{eq:ou_init}
    a^{(x, y, z)}_{\bf k}(0) = \sigma Z.
\end{equation}
For all simulations we set $t_{\rm decay}\equiv100$~Myr.

We consider purely solenoidal driving, so the divergence of the forcing field is cleaned:
\begin{equation}
    {\bf a}_{{\bf k}, \rm div-free} = \left( \textit{\textbf {I}} - \frac{{\bf k}{\bf k}}{|{\bf k}|^2}\right) \cdot {\bf a}_{\bf k},
\end{equation}
where ${\bf a}_{\bf k} = \left(a^{(x)}_{\bf k}, a^{(y)}_{\bf k}, a^{(z)}_{\bf k}\right)$; and \textit{\textbf{I}} is the unit tensor. Finally, the forcing field in real space is obtained by inversely Fourier transforming ${\bf a}_{{\bf k}, \rm div-free}$.

In our simulations, only modes with wavenumber $ k_{\rm min}<|{\bf k}|<k_{\rm max}$ are stirred. Note that the evolution of the Fourier mode does not depend on ${\bf k}$ (Eq.~\ref{eq:ou},\ref{eq:ou_init}); and the divergence cleaning process only introduces dependence on ${\bf e}_{k}={\bf k}/|{\bf k}|$. Thus, the spectrum of the resultant forcing field is a top-hat function from $k_{\rm min}$ to $k_{\rm max}$. We set $k_{\rm min}$ to be close to $k_{\rm max}$ so that the turbulence generated by the spectral forcing approximately has a single driving scale of $l_{\rm drive}=\frac{2\pi}{k_{\rm peak}}$, where $k_{\rm peak}=(k_{\rm min}+k_{\rm max})/2$. 
The turbulence then develops self-consistently, cascading down to scales smaller than $l_{\rm drive}$.

Ideally, the parameter $\epsilon_{\rm inj}$ measures the energy injected by the forcing field per mode, since 
\begin{equation}\label{eq:ou_cov}
    \langle{\bf a}(t)\cdot {\bf v}(t)\rangle_{\bf x} =\left\langle \int_0^t {\bf a}_{\bf k}(t)\cdot {\bf a}_{\bf k}(t') {\rm d} t'\right\rangle_{\bf k} = \sigma^2 \int_0^t e^{-(t-t')/t_{\rm decay}}{\rm d} t',
\end{equation}   
where $\langle\rangle_{\bf x}$ and $\langle\rangle_{\bf k}$ represent averaging over the real and wavenumber space, respectively. Note that the covariance of $a_{\bf k}(t)$ generated by Eq.~\ref{eq:ou} is $\langle a_{\bf k}(t_1)a_{\bf k}(t_2)\rangle=\sigma^2 {\rm exp}(-|t_1-t_2|/t_{\rm decay})$ \citep{Bartosch01}. For $t\gtrsim t_{\rm decay}$, Eq.~\ref{eq:ou_cov} gives $\langle{\bf a}(t)\cdot {\bf v}(t)\rangle_{\bf x} = \sigma^2t_{\rm decay}=\epsilon_{\rm inj}$.
However, Eq.~\ref{eq:ou_cov} holds only when the velocity is all generated by the random forcing. In practice, velocity can be contributed by non-ideal factors in our simulations such as gas outflow and internal gravity waves. Therefore, although we vary the parameter $\epsilon_{\rm inj}$ among the runs to obtain different velocity dispersion, $\epsilon_{\rm inj}$ is only an approximate indicator of the injected energy; and we obtain the actual turbulence energy injection rate by directly measuring $\langle{\bf a}(t)\cdot {\bf v}(t)\rangle_{\bf x}$ in the simulations. 

\subsection{Initial conditions}\label{sec:initialCondition}
Table~\ref{tab:listOfSims} lists all simulations performed and the key parameters. In this section we describe the simulation setups and justify the choices of the parameters. All runs are in 3D, in spherical geometry. 

We set up three runs, \noga, \nogb, and \nogc\  with zero gravity to investigate the turbulent heating in the unstratified regime. The simulation domain is a $(500{\rm kpc})^3$ cube resolved by $320^3$ zones; thus the resolution is $\Delta x\approx1.56\;{\rm kpc}$. In order to test the mixing length theory where the energy flux due to turbulence is proportional to the entropy gradient, we set the initial entropy profiles of these runs to be a power law. The gas electron density and temperature are set to be:
 $n_e(r)=0.1r_{\rm kpc}^{-\alpha}{\rm cm^{-3}}$ and $ T(r)=3r_{\rm kpc}^{\alpha}{\rm keV}$, such that the gas in these no gravity runs is initially isobaric. Thus the power law entropy profile is: 
 \begin{equation}
 K = K_0 r_{\rm kpc}^{\alpha_K} =  T/n_e^{2/3} \approx 14  r_{\rm kpc}^{5\alpha/3}{\rm keV\; cm^2 }.
 \end{equation}
We set $\alpha_K=1/2, 5/6,$ and $4/3$ for the \noga, \nogb\ and \nogc\ run; the density and temperature profiles are determined accordingly, since $\alpha=3 \alpha_K/5$. 
We set $\epsilon_{\rm inj}=8\times10^{-6}{\rm erg\; g^{-1}\; s^{-1}}$ and $l_{\rm drive}=20 \, {\rm kpc}$. 
The resultant turbulence in these different runs has velocity dispersion ranging in $100\sim140{\rm km\, s^{-1}}.$ (this variation arises because the density profiles vary, while the energy injection rate is fixed). 

We perform five runs to study how gravitational stratification affects turbulent heating. 

First, we include one run, \per, with initial gas conditions consistent with the Perseus cluster.  We adopt an analytical fit of the temperature profile based on the observed X-ray surface brightness of Perseus \citep{2003ApJ...590..225C}:
\begin{equation}
    T(r) = 7{\rm keV} \frac{1+(r_{\rm kpc}/71)^{3}}{2.3+(r_{\rm kpc}/71)^3} [1+(r_{\rm kpc}/380)^2]^{-0.23}.
\end{equation}
We consider the gravitational potential due to stars and dark matter, which does not evolve in our simulations. 
The dark matter potential is described by an NFW profile \citep{1996ApJ...462..563N}, with virial radius $r_{\rm vir}=2.44{\rm Mpc}$, virial mass $M_{\rm vir}=8.5\times10^{14}M_\odot$, and the concentration parameter $c=6.81$.
The gravitational acceleration due to stars is described by an analytical fit to the de Vaucouleurs profile of the stellar mass of NGC 1275, the brightest cluster galaxy of Perseus cluster \citep{2006ApJ...638..659M}:
\begin{equation}\label{eq:stellarG}
g_{\rm star}(r) = \left[\frac{r_{\rm kpc}^{0.5378}}{2.853\times10^{-7}} + \frac{r_{\rm kpc}^{1.738}} {1.749\times10^{-6}} \right]^{-1.11}{\rm cm\; s^{-2}}.
\end{equation}
Using the analytical profiles of temperature and gravitational acceleration, we solve the hydrostatic equilibrium equation for the initial gas density profile, which is normalized to match the observed density profile \citep{2006ApJ...638..659M}. The spectral forcing of the \per\ run has the same parameters as in the unstratified runs; and the resultant turbulence is weakly stratified, with the azimuthally-averaged ${\rm Fr}$ ranging from $0.4-0.9$. 

Second, to study the physics of strongly stratified regime, we consider four runs \stra, \strb, \strc, and \strg, among which $\epsilon_{\rm inj}$ and gravitational acceleration are varied to 
explore the parameter space of ${\rm Fr}$.
In order to better resolve the inner ICM region where ${\rm Fr}$ is the smallest, we reduce the size of the simulation domain to $125^3{\rm kpc}^3$ and keep the number of zones unchanged. Thus, the resolution of all the strongly stratified runs is $\Delta x\approx0.39{\rm kpc}$. 
The runs \stra, \strb, and \strc\ have the same gravitational potential as the \per\, runs, but lower energy injection rates $\epsilon_{\rm inj}=6.4\times10^{-7}, 3.2\times10^{-7},$ and $1.2\times10^{-7} {\rm erg\; g^{-1}\; s^{-1}}$, respectively (corresponding to turbulent velocities $u_{\rm rms} = (70,55,43) {\rm km \, s^{-1}}$ at an outer scale of $l_{\rm drive} = 50$ kpc). The run \strg\ has the same $\epsilon_{\rm inj}$ as \strb; and has 10 times stronger gravitational acceleration. The resultant turbulent velocity is $u_{\rm rms}\lesssim70\; {\rm km\; s}^{-1}$ at a turbulent driving scale $l_{\rm drive} = 50\;{\rm kpc}$. This gives: 
\begin{equation}
    {\rm Fr} \approx 0.11 \left(\frac{u_{\rm rms}}{70\;{\rm km\; s}^{-1}}\right)
    \left(\frac{l_{\rm drive}}{50\;{\rm kpc}} \right)^{-1}
    \left(\frac{N_{\rm BV}}{4\times10^{-16}{\rm s}^{-1}}\right), 
\end{equation}
i.e., by construction the strongly stratified runs have ${\rm Fr}\lesssim0.1$. 

For the stratified runs, we use a smoothed broken power law model for the initial gas entropy profiles \citep{2018ApJ...862...39B}:
\begin{equation}\label{eq:univEnt}
    K(r) = K_0 \left(\frac{r}{r_b}\right)^{p_1} \left[ \frac{1}{2}+\frac{1}{2} \left( \frac{r}{r_b}\right)^{1/\Delta} \right]^{(p_2-p_1)\Delta},
\end{equation}
where $p_1$ and $p_2$ are the power-law slopes below and above the break radius, $r_{b}$, respectively; and $\Delta$ is the parameter controlling the smoothness of the change of slope. We set $\Delta=0.2$ such that slope change approximately occurs within $0.6 - 1.6 \, r_b$. For \stra, \strb, and \strc, we adopt $p_1=0.65$, $p_2=1.02$, and $r_{b}=16.8{\rm kpc}$, which are obtained by fitting the universal entropy profile observed in cool-core clusters \citep{2018ApJ...862...39B}.  Then the gas density and temperature profiles are obtained by solving for hydrostatic equilibrium with the given entropy profile. For the \strg\ run, the gas entropy and the gravitational acceleration are 10 times larger, which results in 10 times higher gas temperature, but the same initial gas density. This modified profile is constructed to provide a clean test of physics in the low {\rm Fr} number regime.

\begin{figure*}
	\includegraphics[width=\textwidth]{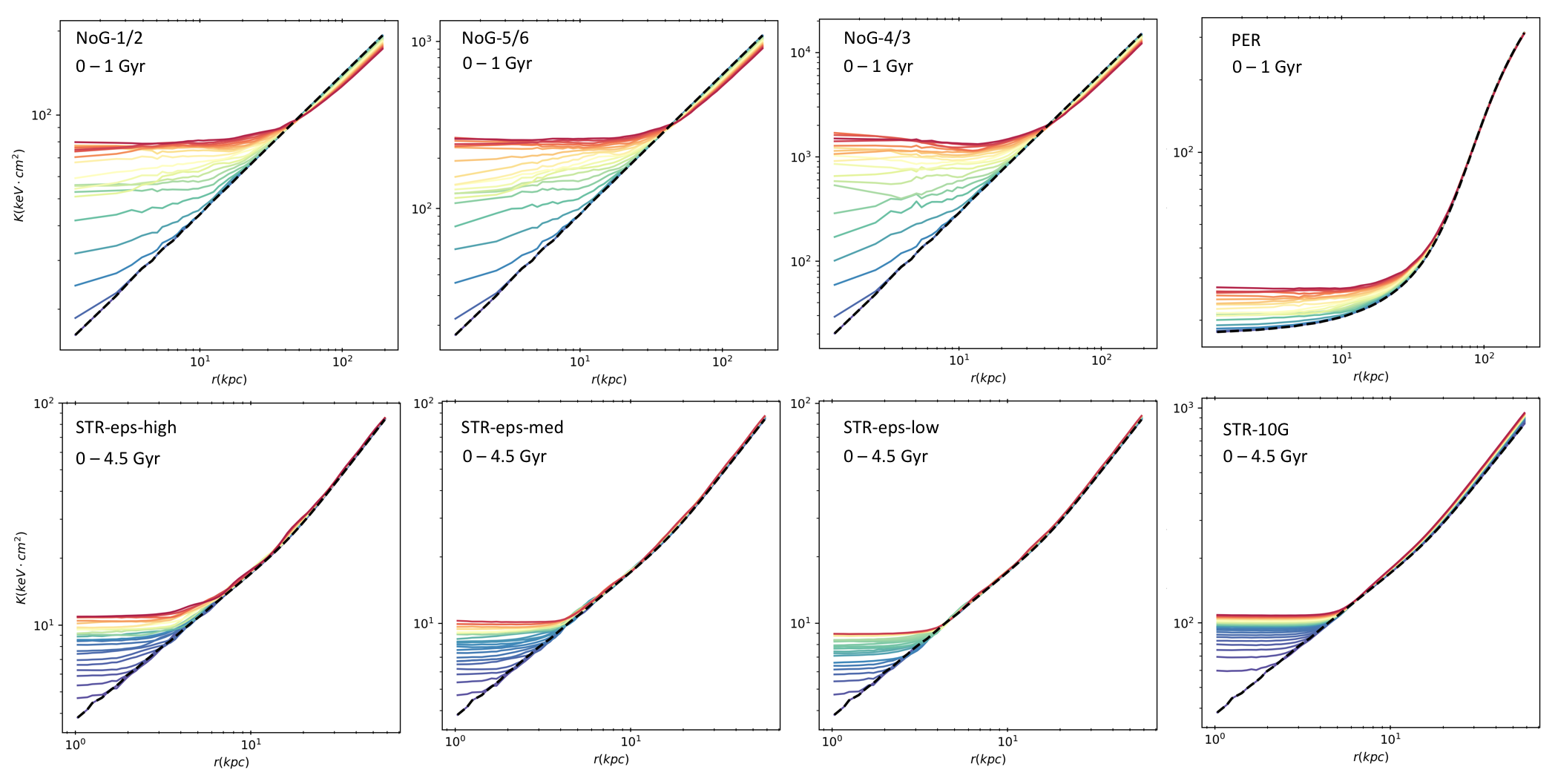}
    \caption{Time evolution of radial entropy profiles. The order of the panels are the same as that of Fig.~\ref{fig:sigAll}. Simulation time is encoded by the color, from purple being the earliest to red being the latest. The initial conditions are highlighted by the dashed black lines. For the no-gravity and weakly-stratified cases (top row), the profiles are drawn every 50~Myr for $t=0-1~{\rm Gyr}$; and for the strongly-stratified cases (bottom row), the profiles are drawn every 200~Myr for $t=0-4.5~{\rm Gyr}$.
    In all cases, an isentropic core grows in radius due to convective heating; and the region outside the core generally remains adiabatic. By comparing the top and bottom rows, it is clear that
    strong stratification results in smaller cores, which develop over much longer time scales.}  
    \label{fig:entAll}
\end{figure*}

\section{Main Results}
\label{sec:results} 
\subsection{General evolution}

As the simulation proceeds, the spectral forcing scheme drives turbulence in the halo. As shown by Fig.~\ref{fig:sigAll}, the rms turbulent velocity $u_{\rm rms}$ increases with time and reaches a plateau after an initial rise. 
The plateau marks the stage where the turbulent energy injection rate is balanced by the cascade rate of turbulence to smaller scales. 
The time needed to reach the stable state, $t_{\rm int}$, lasts for several eddy turn-over time, $t_{\rm int} = nt_{\rm eddy}$. The upper ticks of each panel in Fig.~\ref{fig:sigAll} denote the simulation time scaled by $t_{\rm eddy}$. $n$ gets larger for stronger stratification: for unstratified and weakly-stratified runs (top panels), $n\sim1.5$, while $n$ increases from $\sim1.5$ to $\sim3$ for the strongly-stratified runs in the order of increasing strength of stratification (the lower panels, from left to right). The correlation between $n$ and the strength of stratification is consistent with the fact that cascade rates to smaller scales are weaker with stronger stratification, and thus turbulence takes a higher number of eddy turnover times to saturate. 

Fig.~\ref{fig:entAll} shows the snapshots of gas entropy radial profiles. Simulation time of each snapshot is color-coded such that redder colors represent later times. The initial conditions are highlighted with the black-dashed lines. All runs exhibit similar evolution: turbulence creates an isentropic core growing in radius; this central core entropy increases with time. Outside the growing core, gas flows out of the boundary and the halo expands adiabatically.  
The outer entropy profile does not change significantly with time. 

The rise of entropy in the central region results from entropy influx due to turbulent diffusion in the absence of radiative cooling. Note that there are no buoyant restoring forces in the isentropic core; the Froude number diverges there, and this could bias our analysis. For the diffusion analysis, the isentropic region is excluded: the diffusive flux in Fig. 5 starts from about 20 kpc for NoG and PER runs, and from about 5 kpc for STR runs. Compared with the isotropic and weakly-stratified cases (top panels), the isentropic cores found in the strongly stratified cases (bottom panels) grow more slowly and are restricted to a smaller region (the inner $\sim5\;{\rm kpc}$) at the end of the simulations. This reflects the suppression of turbulent diffusion in a strongly stratified medium. For dissipation analysis, the isentropic core is only a small fraction of the analyzed volume, and it has negligible impact on our results. Note that in high resolution observations, power-law entropy profiles, rather than isentropic cores, are seen at the centers of clusters \citep{Babyk2019}.  

\begin{figure}
	\includegraphics[width=\columnwidth]{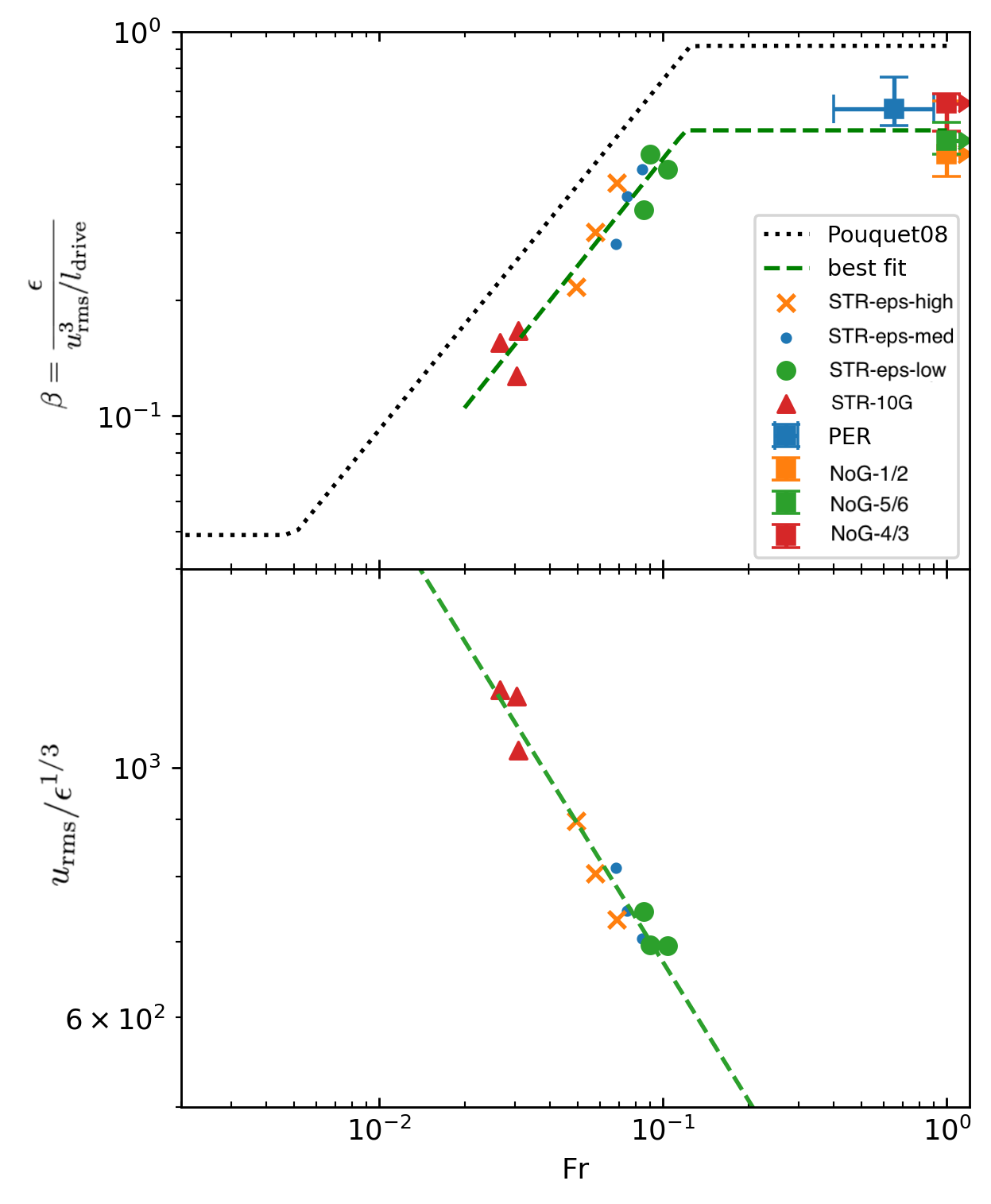}
    \caption{{\it Top panel}: the Froude number ($\rm Fr$) scaling of turbulent dissipation efficiency ($\beta = \epsilon/(u_{\rm rms}^3/l_{\rm drive})$), the ratio between the actual turbulent dissipation rate $\epsilon$ and the Kolmogorov rate $u_{\rm rms}^3/l_{\rm drive}$.
    In the strongly-stratified cases, time- and volume-averaged $\beta$ and $\rm Fr$ are calculated in each of the three analyzed radial layers. The results are shown as the data points without bars.
    For the no-gravity and weakly-stratified cases, which are not split up by radial shells, the range of $\beta$ over the domain is shown by the vertical bar. The range of time-averaged $\rm Fr$ for the \per\ case is shown as the horizontal bar; and the right arrows on the data points of no-gravity cases representing $\rm Fr=\infty$ in these cases. The green dashed line shows the best fit broken power law (Eq.~\ref{eq:betaFr}). The $\beta-{\rm Fr}$ relation obtained by \citet{pouquet18} is shown as the black dashed line.
    {\it Bottom panel}: averaged values of ${\rm Fr}$ and $u_{\rm rms}/\epsilon^{1/3}$ in the radial shells of the strongly-stratified cases. The green dashed line shows the best fit power law. 
    Both $\beta$ and $u_{\rm rms}/\epsilon^{1/3}$ tightly relate to ${\rm Fr}$, and the power indices in the strongly-stratified regime agree well with that predicted by the wave-turbulence model (Eq.~\ref{eq:vFr}, Eq.~\ref{eqn:tau_cas_Fr}).
    }
    \label{fig:dissMain}
\end{figure}

\begin{figure}
	\includegraphics[width=\columnwidth]{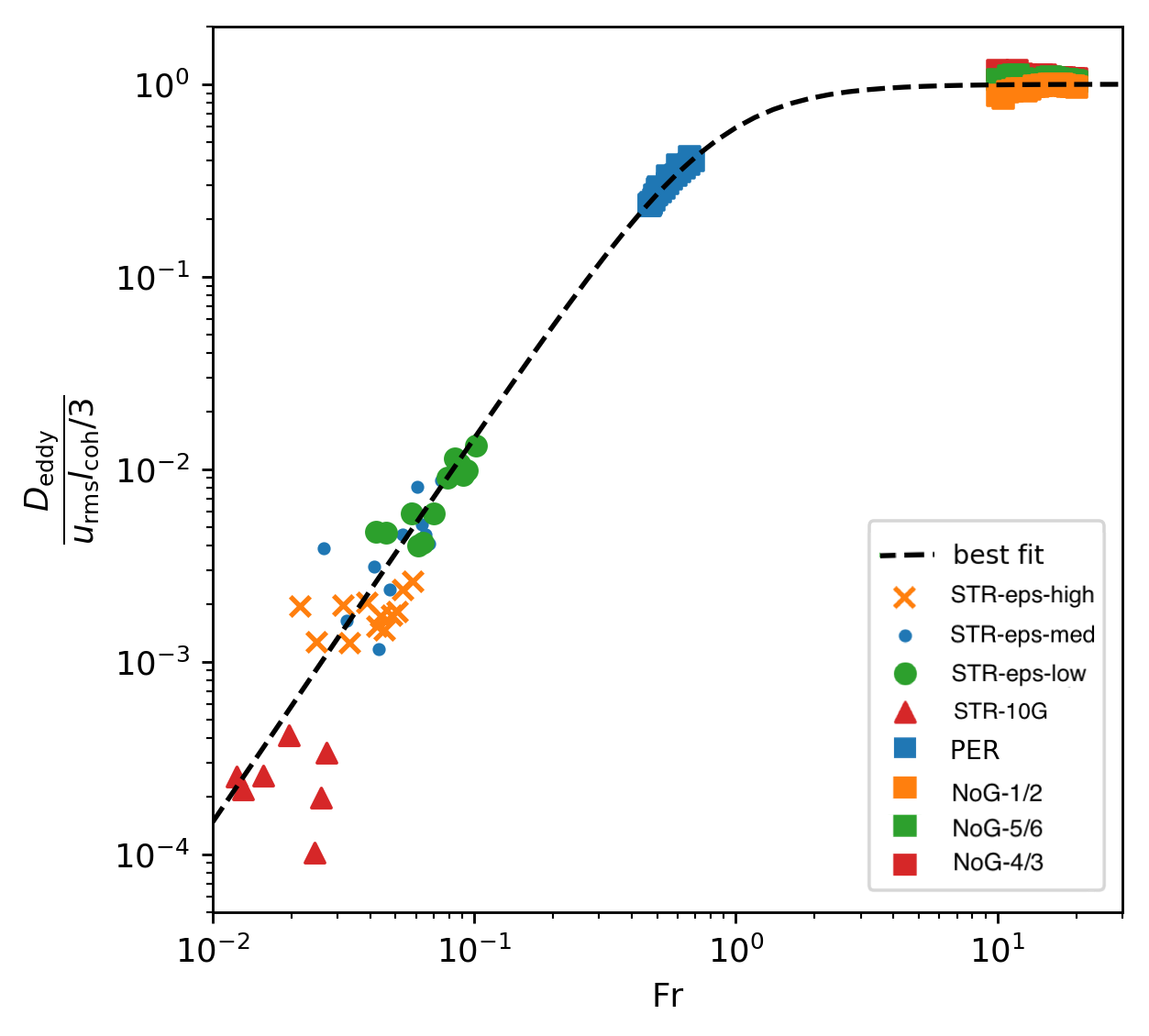}
    \caption{Suppression of turbulent diffusion coefficient as a function of ${\rm Fr}$. The black dashed line is the analytical model (equation \ref{eq:stratDiff}) fitted by the data from the \per run (the blue squares). For convenience we plot the data from the no-gravity run at the right end (${\rm Fr\sim15}$). The plotted symbols retain the same meaning as Fig \ref{fig:dissMain}.   }
    \label{fig:deddyFr}
\end{figure}

\begin{figure*}
     \centering
    \includegraphics[width=\textwidth]{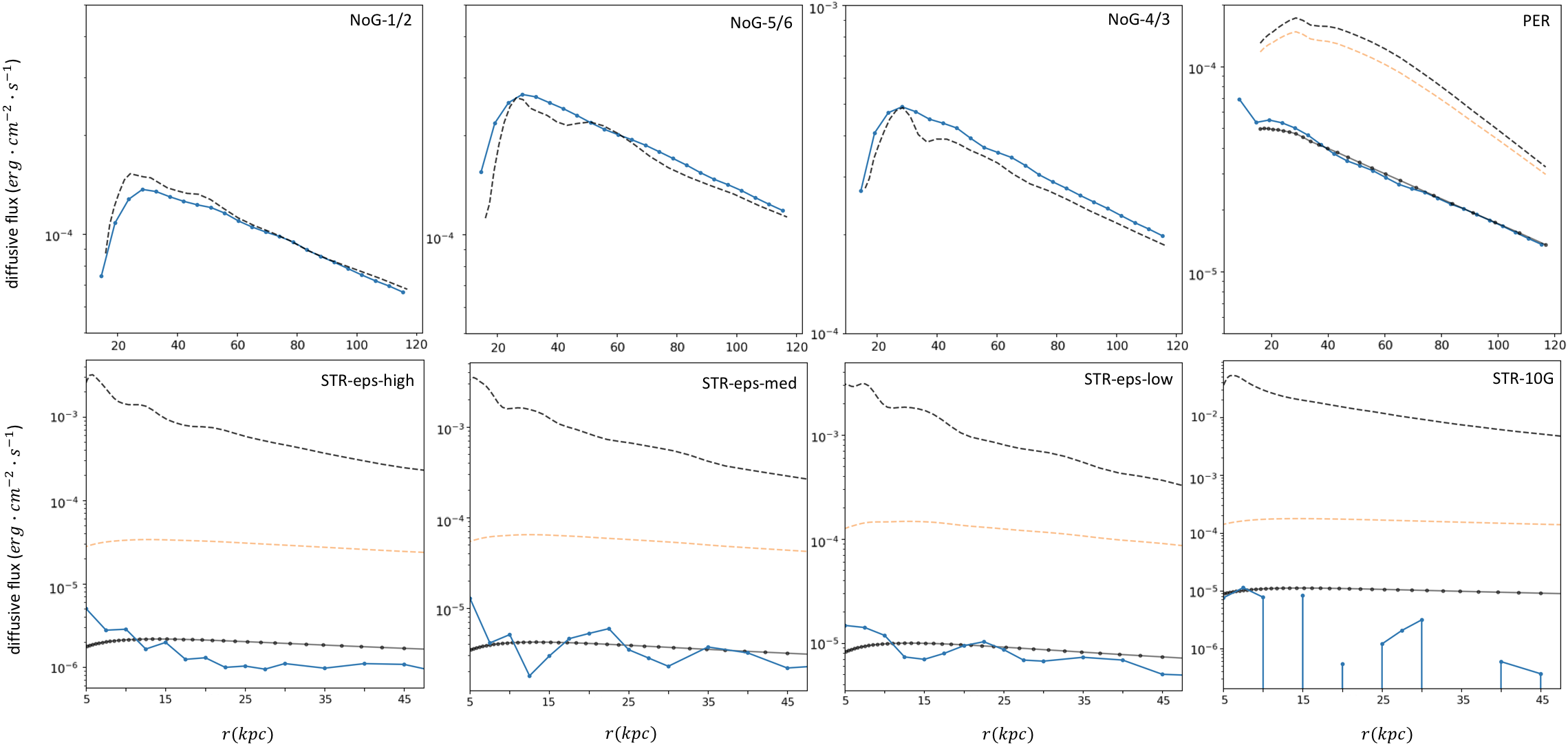}  
       \caption[]{Averaged radial profiles of the convective energy flux due to turbulent diffusion. The dashed grey lines correspond to the flux predicted from the standard mixing length theory, where the correction for gravitational stratification is not considered; the blue lines correspond to the actual convective flux measured from the simulations; the dotted black line corresponds to the prediction from the mixing length theory corrected for stratification with the best fit value of the free parameter $c_1\approx0.68$ in Eq.~\ref{eq:stratDiff}; and the red dashed lines correspond to the stratified mixing length model with the $c_1$ value reported in \citet{weinstock81}. All profiles are averaged over the simulation time when the turbulence has reached a steady state.
    }
     \label{fig:diffFluxAll}
\end{figure*}

\begin{figure}
	\includegraphics[width=.95\columnwidth]{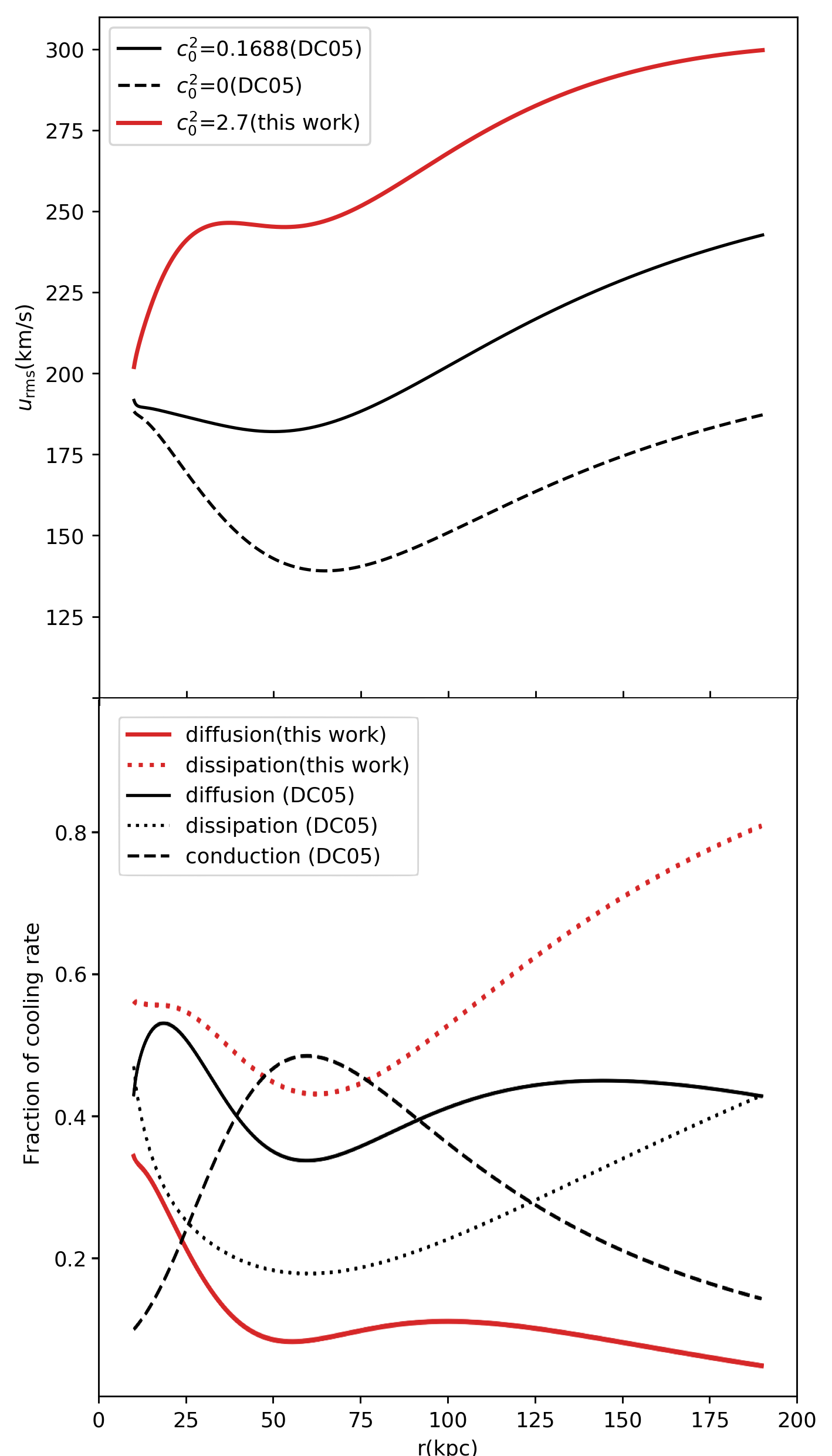}
    \caption{Top panel: velocity profiles of a galaxy cluster where heating due to turbulent dissipation, diffusion and thermal conduction balance radiative cooling, assuming different models of turbulent diffusion. We reproduce the results for cluster A1795 in DC05, shown as the black line. DC05 adopts $c_1=0.042$ (as well as $c_0^2=0.1688$) for the turbulent diffusion coefficient (Eq.~\ref{eq:stratDiff}). The red line shows the resultant velocity profile with $c_1=0.68$ calibrated by this work.
    Bottom panel: relative contribution of different heating sources to balance radiative cooling. The results for cluster A1795 in DC05 ($c_1=0.042$) are reproduced, shown as the black lines, where the solid, dotted, and dashed lines correspond to the contribution from diffusion, dissipation, and thermal conduction, respectively. The red lines show the contribution from diffusion and dissipation in the case of $c_1=0.68$, which is calibrated from our numerical simulations. 
        } 
    \label{fig:dc05}
\end{figure}

\begin{figure}
	\includegraphics[width=\columnwidth]{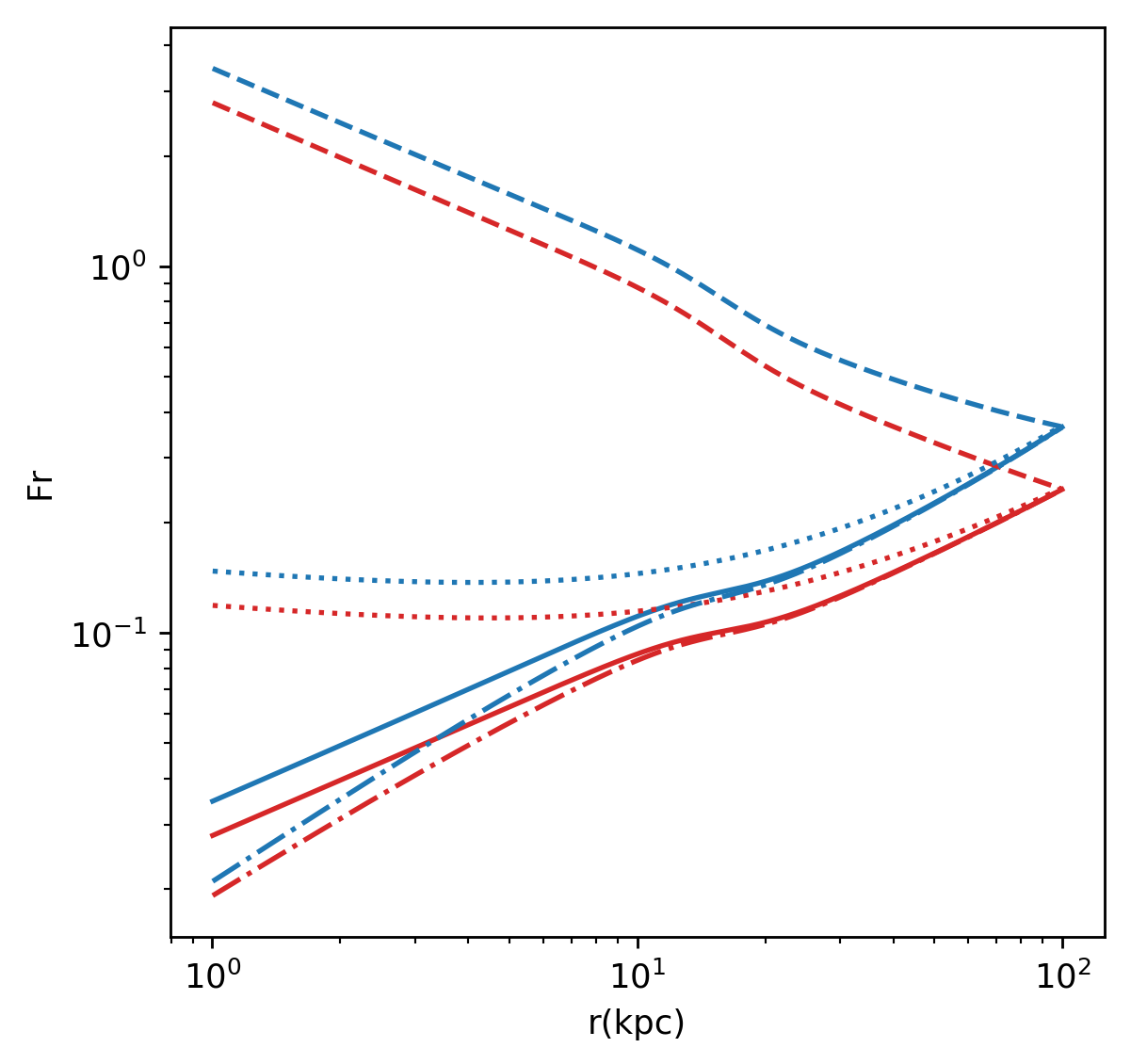}
    \caption{
        Radial profiles of ${\rm Fr}$ with different assumptions about driving scale, entropy gradient and gravitational potential. The rms turbulent velocity is set to be $u_{\rm rms} = 200$\,km/s. 
        The solid blue and red lines correspond to a dark matter potential described by the NFW profile with $M_{\rm 500}=10^{14}, 10^{15}M_{\odot}$, respectively; gas entropy follows the universal profile (Eq.\ref{eq:univEnt}); and the turbulent driving scale is $l_{\rm drive} =100$\, kpc. Since ${\rm Fr} \propto u_{\rm rms}/l_{\rm drive}$, the results here for other (spatially constant) values $u_{\rm rms}, l_{\rm drive}$ can be found by simple rescaling
        Compared with the solid lines, other blue \& red line pairs have one of the following differences: 1) the driving scale equals to $r$ (dashed lines); 2) the gas entropy profile has a flat core (dotted lines); 3) an additional gravitational potential from the stellar mass (using stellar potential of NGC1275 as an example) is included (dotted dash lines).
    }
    \label{fig:FrProfile}
\end{figure}

\subsection{Turbulent dissipation}
{\bf Strongly stratified cases.} For the strongly stratified cases, we calculate the turbulent dissipation rate in three different radial shells: $20<r_{\rm kpc}<30$, $30<r_{\rm kpc}<40$, and $40<r_{\rm kpc}<50$. 
The actual turbulent dissipation rate is estimated by the energy injection rate of the turbulent forcing, i.e., $\epsilon_{\rm diss} \approx \langle {\bf a\cdot u} \rangle$, where $\langle\rangle$ stands for averaging over both time and volume.
Thus the deviation of turbulence dissipation rate from the prediction of Kolmogorov-like turbulence, $\beta$ is: 
\begin{equation}
    \beta = \frac{ \langle {\bf a\cdot u} \rangle}{\langle u_{\rm rms}^3/l_{\rm drive}\rangle},
\end{equation}
where ${\bf a\cdot u}$ and $u_{\rm rms}^3/l_{\rm drive}$ are evaluated by averaging over the stable 
period of simulation and over the volume of each radial shell. 
There is a tight correlation between $\beta$ and the averaged ${\rm Fr}$ in the strongly-stratified cases, as demonstrated in the upper panel of Fig.\ref{fig:dissMain}. The best fit (green dashed line) model suggests an approximate linear relation in this regime, $\beta \sim {\rm Fr}$. $\beta$ can be interpreted as the ratio between the eddy turn-over time and the turbulent dissipation time, since:
\begin{equation}
    \beta \sim \frac{\epsilon_{\rm diss}}{u^2/ (l/u) }\sim \frac{u^{2}/\tau_{\rm diss}}{u^2/\tau_{\rm eddy}}\sim\frac{\tau_{\rm eddy}}{\tau_{\rm diss}}.
\end{equation}
As discussed in section \ref{sec:stratTurb}, $\beta\propto{\rm Fr}$ is expected for the wave turbulence in a gravitationally stratified medium. In general, turbulent cascade and dissipation rates are lower in the wave turbulence regime; and stronger stratification (lower $\rm Fr$) leads to lower turbulence dissipation rates for a given $u_{\rm rms}$. 

Additionally, the turbulent velocity dispersion is consistent with the wave turbulence picture. Within one individual run, the turbulent energy injection rate, $\epsilon_{\rm diss}$, is approximately unchanged among the three radial shells. For a given injection rate, the resultant velocity dispersion increases as ${\rm Fr}$ falls due to the increase in the turbulent cascade time predicted by wave turbulence theory (section \ref{sec:stratTurb}). Inner radial shells (which have lower Fr) have larger velocity dispersion, as demonstrated in the bottom row of Fig.~\ref{fig:sigAll}. 
Quantitative analysis also show consistency. 
The bottom panel of Fig.~\ref{fig:dissMain} demonstrates the scaling between $\frac{u_{\rm rms}}{\epsilon_{\rm diss}^{1/3}}$ and ${\rm Fr}$. The data reveals a tight power-law relation (the green dashed line): 
\begin{equation}
\frac{u_{\rm rms}}{\epsilon_{\rm diss}^{1/3}}\propto {\rm Fr}^{-0.41}.
\end{equation}
The power-law scaling has a slope very close to that predicted from wave turbulence theory (Eq.~\ref{eq:vFr}).

{\bf Unstratified and weakly stratified cases}
In the unstratified and weakly stratified runs, turbulence is more isotropic; we find the turbulent dissipation is consistent with the \citet{kolmogorov41} prediction. In Fig.~\ref{fig:dissMain} we show $\beta$ for each run averaged over time and the entire volume of the simulation as square data points. We denote the minimum and maximum value of time-averaged $\beta(r)$ as vertical error bars. For the data point of the \per\ run, the range of ${\rm Fr}(r)$ is shown as the horizontal error bar. In the unstratified cases, ${\rm Fr}\rightarrow\infty$, which is represented by the right arrows attached to the data points. The radial variation of $\beta$ throughout the entire simulation domain is small. Variation of $\beta(r)$ in \per\ is slightly larger, but is still very small given the range of ${\rm Fr(r)}$. Therefore, $\beta$ in the isobaric and weakly stratified cases can be treated as a constant. 
This also implies that density stratification in its own right does not affect the dissipation rate.

Combining the results of all runs we fit the $\beta-{\rm Fr}$ data points with a  piecewise function:
\begin{equation}\label{eq:betaFr}
    \beta = \begin{cases}
      3.9~{\rm Fr}^{0.92}  & 0.03<{\rm Fr} <0.12 \\
      0.55                 & {\rm Fr} \geq 0.12
\end{cases}.
\end{equation}

The scaling relation is in good agreement with that in \citet{pouquet18} (the black dashed line\footnote{Note that \citet{pouquet18} use the integral length $l_{\rm int}\approx \frac{1}{2}l_{\rm drive}$ in $\beta$ and ${\rm Fr}$. This makes $\beta$ (${\rm Fr}$) defined in \citet{pouquet18} two times smaller (larger) than that defined in this paper.} in Fig.~\ref{fig:dissMain}) in terms of the power-law index in $0.03<{\rm Fr} <0.12$ and the critical value of ${\rm Fr}$ above which $\beta$ saturates. 
There is a difference in normalization, which is less than a factor of 2. The different normalization may be due to different fractions of turbulent energy converted to gravitational potential energy, in the respective plane-parallel and spherical set-ups. 

\begin{figure*}
	\includegraphics[width=\textwidth]{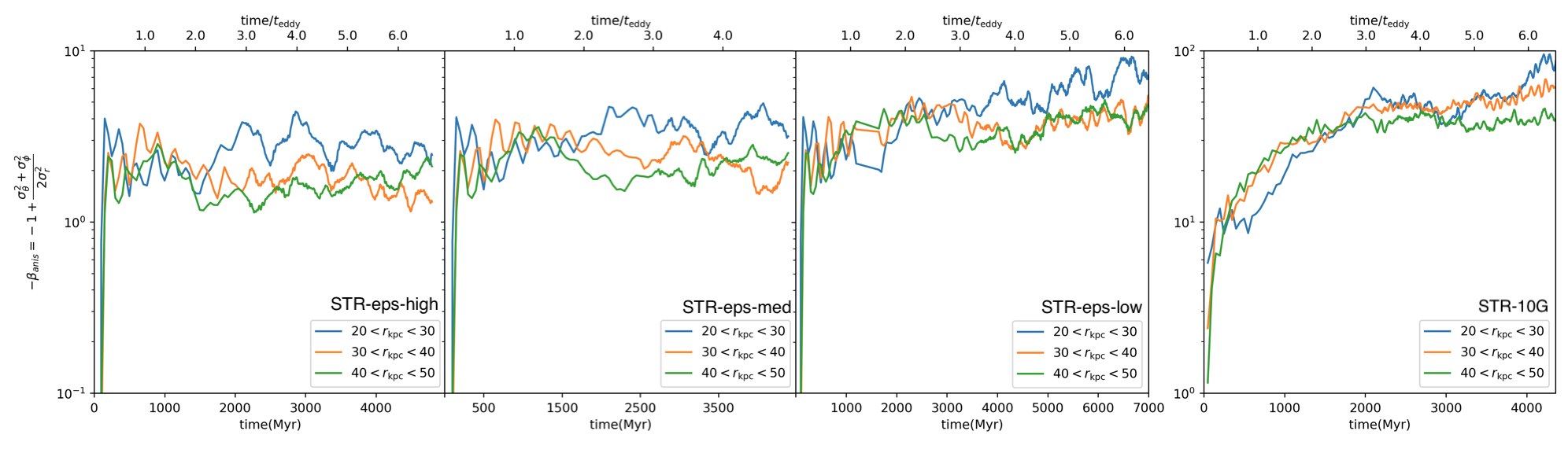}
    \caption{Time evolution of anisotropy parameter, $\beta_{\rm anis}$ in the strongly stratified runs. The velocity anisotropy is defined to be $\beta_{\rm anis} = 1-\frac{\sigma_{\theta}^2+\sigma_{\phi}^2}{2\sigma_r^2}$, where $\sigma_{\theta}$, $\sigma_{\phi}$, and $\sigma_{r}$ are the three spherical components of velocity dispersion. For visibility, $-\beta_{\rm anis}$ is plotted. $-\beta_{\rm anis}$ of the inner shells is larger than that of the outer in each run, indicating more tangentially-biased turbulent velocity hence stronger stratification in the inner region.}
    \label{fig:ansBetaAll}
\end{figure*}

\subsection{Turbulent heat diffusion}

We calculate the energy flux due to turbulent diffusion in the simulated halos using the convective flux of gas enthalpy (${\mathbf F}_{\rm conv}$):
    \begin{equation}\label{eq:fconv}
         {\mathbf F}_{\rm conv}(r) = F_{\rm conv}(r){\mathbf e}_r =  \frac{\gamma}{\gamma-1}k_B\left( \left<nu_rT\right>-\left<nu_r\right>\left<T\right>\right){\mathbf e}_r,
    \end{equation}
    where $k_B$ is the Boltzmann constant, $n$ is the gas number density, $u_r$ is the radial velocity, and $<>$ represents averaging over the $4\pi$ solid angle, e.g.,
    $\left<T\right> = \dfrac{1}{4\pi}\int T(r, \Omega) {\rm d}\Omega $. The convective heat flux 
    is obtained by subtracting the heat flux due to mass inflow/outflow
    from the total heat flux. The actual heating rate due to the convective flux ($\dot{e}_{\rm conv}(r)$) is the negative divergence of the convective flux. Considering spherical symmetry, $\dot{e}_{\rm conv}(r)$ is:
\begin{equation}
    \dot{e}_{\rm conv}(r) = -\nabla\cdot{\mathbf F}_{\rm conv}(r) = -\frac{1}{r^2}\frac{\partial (r^2F_{\rm conv})}{ \partial r}.
\end{equation}

Then we compare the calculated turbulent heat diffusion with that prediced by mixing length theory (\citealt{kim03}; DC05), where the diffusive energy flux is proportional to the entropy gradient:

\begin{equation}\label{eq:mlt}
    {\mathbf F}_{\rm MLT}(r) = F_{\rm MLT}(r) \mathbf{e}_r = -D_{\rm eddy} \rho(r)T(r) \frac{\partial s(r)}{\partial r} \mathbf{e}_r,
\end{equation}
where $s(r)=c_v {\rm ln}\left(\frac{P(r)}{\rho(r)^\gamma}\right)$ is the specific gas entropy, $c_v=\frac{k_B}{\mu m_p(\gamma-1)}$ is the heat capacity at constant volume per unit mass, $m_p$ is proton mass,  $\mu\approx0.6$ is the mean molecular weight of the plasma. $D_{\rm eddy}$ is the turbulent diffusion coefficient.
 Fig.~\ref{fig:deddyFr} shows averaged $\frac{D_{\rm eddy}}{u_{\rm rms} l_{\rm coh}/3}$ as a function of the average ${\rm Fr}$ in the radial bins of each run, where $l_{\rm coh}\approx\frac{1}{4}l_{\rm drive}$ is the coherence length\footnote{The relationship between $l_{\rm coh}, l_{\rm drive}$ comes from integrating over the power spectrum of turbulence \citep{tennekes72}.}; and $D_{\rm eddy}$ is estimated by $D_{\rm eddy} = \frac{F_{\rm conv}}{\rho T\partial s/\partial r}$. Averaging is performed over the volume of radial bins and time during which turbulence has reached a stable state and does not evolve systematically. 

 Fig.~\ref{fig:deddyFr} demonstrates that $D_{\rm eddy}$ is remarkably well described by Eq.\ref{eq:stratDiff} from no-stratified to strongly stratified cases. The black dashed line shows the best-fit analytical expression for the correction factor due to stratification,
 \begin{equation}
 \frac{D_{\rm eddy}}{u_{\rm rms} l_{\rm coh}/3}=\frac{1}{1+0.68{\rm Fr}^{-2}}. 
\end{equation} 
We describe the turbulent diffusion in detail below for the isotropic, weakly stratified and strongly stratified cases respectively.

{\bf Unstratified and weakly stratified cases}
The standard mixing length theory ignores gravitational stratification, so $Fr\rightarrow\infty$; $D_{\rm eddy}= u_{\rm rms}l_{\rm cho}/3$; and the correction factor reduces to unity, as shown by the orange, red and green squares in Fig.~\ref{fig:deddyFr}.
Therefore, the results for the isotropic cases (i.e., runs \noga, \nogb, and \nogc), where gravity is excluded, demonstrate that the heating rate due to turbulent diffusion agrees  with the standard mixing theory very well. 
As shown in the top left three panels in Fig.~\ref{fig:diffFluxAll}, the diffusive heat flux from the standard mixing length model (the grey dashed lines) is in good agreement with the heat flux due to turbulent diffusion measured from simulations (the blue lines). 
On the other hand, in \per\ run, where weak stratification is present, the prediction from standard mixing length theory significantly deviates from the actual diffusion heating rate. This is demonstrated by the offset between the dashed grey line (representing the prediction from standard mixing length theory) and the blue line (representing the values measured from the simulation) in the top right panel in Fig.~\ref{fig:diffFluxAll}. However, the mixing length model corrected for the gravitational stratification, i.e., $D_{\rm eddy}\sim\frac{ul}{1+c_1{\rm Fr}^{-2}}$, (dotted solid black line) can accurately describe the turbulent mixing heating rate directly measured from the numerical simulations. The best fit requires 
\begin{equation}
    c_1\approx0.68,
\end{equation} 
which is a factor of 16 larger than the original value calibrated in \citet{weinstock81} for the conditions appropriate for the Earth's atmosphere. 
We have checked carefully that this difference does not arise due to a difference in the definition of coherence length and driving scale. In particular, for a given driving scale $l_{\rm drive}$, we reproduce the analytic results of \citet{dennis05}, given their equations (see below). However, these equations do not agree with our numerical simulations. 

{\bf Strongly stratified cases} For strongly stratified cases, discrepancies between the standard mixing length theory and the stratified model are much larger, as shown by the grey dashed lines and dotted solid black lines in the bottom panels of Fig.~\ref{fig:diffFluxAll}. This is because ${\rm Fr}\ll1$ and $D_{\rm eddy}\sim\frac{ul}{1+c_1{\rm Fr}^{-2}}\sim ul{\rm Fr}^2$. Namely, the convective flux due to turbulent diffusion is suppressed by a factor of ${\rm Fr}^2$, which corresponds to 2 to 3 orders of magnitude in the strongly-stratified cases. The actual diffusive flux in simulations (dotted blue lines in the bottom panels of Fig.~\ref{fig:diffFluxAll}) is broadly consistent with the prediction of the stratified model, though the measured heating rate has large scatter with negative values. In a strongly-stratified medium, the radial motions are dominated by gravity waves and can be approximated as oscillations with a time scale $\tau_{\rm BV}\sim N_{\rm BV}^{-1}$. However, as shown in section \ref{sec:stratTurb}, the non-linear time $\tau_{\rm NL}\sim \tau_{\rm eddy}{\rm Fr}^{-1}\sim \tau_{\rm BV}{\rm Fr}^{-2}\gg\tau_{\rm BV}$. $\tau_{\rm NL}$ is the timescale over which coherent motions are lost and hence it indicates the time for stochastic motions to sufficiently mix the gas. Therefore, the dominant mode of radial motions almost leaves the gas entropy unchanged, as the oscillation is too swift to allow the displaced gas parcel to mix with the ambient medium. Consequently, the actual convective flux due to turbulent diffusion is swamped by highly fluctuating signals with zero mean; and this causes the large scatter in the averaged diffusion heating rate in the strongly-stratified cases.

\subsection{Relative contributions of turbulent diffusion and dissipation}
Our simulations suggest turbulent diffusion is considerably more suppressed by gravitational stratification than previously thought.
We find this implies that higher velocities are needed to offset cooling. Compared to turbulent dissipation, where dissipation rates $\epsilon \propto {\rm Fr}$ for ${\rm Fr} \lsim 0.1$, suppression of turbulent diffusion is stronger ($D \propto {\rm Fr}^{2}$, and, importantly, sets in at much higher Froude number ${\rm Fr} \lsim 1$). 
Thus, when stratification is strong, turbulent dissipation is generally more important than turbulent diffusion in heating the gas. 

DC05 calculate the velocity profiles of turbulence with which turbulent dissipation and diffusion can balance the radiative cooling together with the thermal conduction. The resultant velocity profile and relative contribution by dissipation and diffusion for cluster A1795 are reproduced. DC05 adopts $c_1=0.042$ (as well as $c_0^2=0.1688$) for the turbulent diffusion coefficient.
We follow their calculation with the same value of $c_0^2$, but with a much larger $c_1=0.68$, which is calibrated by our simulations. 
The top panel of Fig.~\ref{fig:dc05} shows the resultant velocity profiles with $c_1=0.042$ (black line) and $c_1=0.68$ (red line). The stronger suppression of turbulent heating in our model results in a $20\%\sim25\%$ increase of the velocity profiles. 
The bottom panel of Fig.~\ref{fig:dc05} demonstrates the fraction of cooling rate balanced by diffusion (solid lines), dissipation (dotted lines), and thermal conduction (dashed line). The results of DC05 ($c_1=0.042$) are shown in black; and those of our models ($c_1=0.68$) are shown in red.
Adopting $c_1=0.68$ suppresses the turbulent diffusion heating rate; thus, in order to achieve thermal equilibrium the relative contribution of turbulent dissipation has to increase (which requires larger turbulent velocities). Compared to previous models, the relative importance of turbulent diffusion and dissipation is switched in our model.

\subsection{Typical values of Fr in ICM}

To investigate the typical degree of gravitational stratification in ICM, we investigate the radial profile of Fr with different assumptions about driving scale, entropy gradient, and gravitational potential.  
As the fiducial model, 
we consider $l_{\rm drive}\equiv100\,{\rm kpc}$, the gas entropy, $K(r)$ described by the universal profile (Eq.~\ref{eq:univEnt}), and rms turbulent velocity $u_{\rm rms} =200\,$km s$^{-1}$. 
Since ${\rm Fr} \propto u_{\rm rms}/l_{\rm drive}$, the results here for other (spatially constant) values $u_{\rm rms}, l_{\rm drive}$ can be found by simple rescaling. For gravitational potential, we employ the NFW profiles with $M_{500}=10^{14}$ and $10^{15}M_\odot$, respectively, which generally brackets the halo mass range of galaxy clusters. The concentration parameters of the NFW profiles, $c_{500}$ are set according to the $c_{500} - M_{500}$ relation from simulations by \citet{2004A&A...416..853D}. The adopted $c_{500} - M_{500}$ relation is in agreement with observations within $\sim2\sigma$ scatter \citep{vikhlinin06}. The resultant ${\rm Fr}(r)$ is shown as solid lines in Fig.~\ref{fig:FrProfile}, where the blue line corresponds to $M_{500}=10^{14}M_\odot$; and the red line represents the model with $M_{500}=10^{15}M_\odot$. It turns out that Fr$(r)$ is not sensitive to the cluster halo mass: the change of Fr$(r)$ is less than a factor of 2 for an order of magnitude difference of halo mass.

We then alter $l_{\rm drive}$, $K(r)$, and the gravitational potential based on the fiducial model to see how Fr$(r)$ is affected. 
First, for a driving scale that is proportional to the distance from the ICM center $l_{\rm drive} \propto r$, the Fr$(r)$ profile (dashed lines in Fig.~\ref{fig:FrProfile}) is distinct from that with constant $l_{\rm drive}$. While this radial dependence of $l_{\rm drive}$ was assumed by previous works \citep[e.g.,][]{kim03, dennis05}, our simulations are inconsistent with such assumptions (section~\ref{sec:drivingScale}).
Second, using an entropy profile with a flat core increases the central Fr by $\sim0.5$~dex (the dotted lines). Early observations found the general existence of isentropic cores in the hot gaseous halos of galaxy clusters \citep[e.g.,][]{2009ApJS..182...12C}, where the entropy profile is given by 
\begin{equation}\label{eq:unviEntOld}
    K(r) = K_0 + K_1r_{\rm kpc}^{\alpha}.
\end{equation}
At outer radii, entropy follows a power law $\propto r^{\alpha}$ and approaches a constant value, $K_0$ at the center. We calculate Fr$(r)$ with the best fit model reported by \citet{2009ApJS..182...12C}, $K(r)=17.5 {\rm keV\cdot cm^2}+ 148\left(r/100{\rm kpc}\right)^{1.21} {\rm keV\cdot cm^2}$.  
However, more recent works 
show that the flattening of inner entropy is a resolution effect \citep{2014MNRAS.438.2341P, 2017ApJ...851...66H, 2018ApJ...862...39B}.
Our calculation suggests the importance of constraining entropy profiles in estimating Fr. Finally, we consider whether including the gravitational potential from the stellar mass has a large impact on Fr$(r)$. We use the stellar potential of NGC1275 as an example (Eq.~\ref{eq:stellarG}). The resultant Fr$(r)$ is shown as the dotted dash line. Only a small change in Fr$(r)$ is caused by including the stellar potential, even though stellar mass dominates over the dark matter mass at small radii. 

In summary, the value of Fr is 
mostly affected by the gas entropy profile, the driving scale, and the amplitude of turbulent fluctuations, while uncertainties in gravitational potential do not affect Fr too much. 

\subsection{Estimating the driving scale}\label{sec:drivingScale}
In this work, we adopt a single peak scale of the spectral forcing, which corresponds to a constant $l_{\rm drive}$ in each run. We can test this argument by examining our simulations directly, by looking at how velocity anisotropy scales with radius.
The constant $l_{\rm drive}$ causes Fr to be an increasing function of radius, since gravity is stronger (and hence the Brunt-Väisälä frequency raises) in inner regions. As stratification becomes stronger (Fr falls), turbulence motions should become more tangentially biased, due to strong buoyant restoring forces which limit motion in the radial direction. Fig.~\ref{fig:ansBetaAll} shows the velocity anisotropy parameter ($\beta_{\rm anis}$) as a function of time, where \begin{equation} 
\beta_{\rm anis} = 1-\frac{\sigma_{\theta}^2+\sigma_{\phi}^2}{2\sigma_r^2}. 
\end{equation} 
In each run, turbulent velocity is more tangentially-biased (more negative $\beta_{\rm anis}$) in the inner shells, indicating inner regions are more strongly stratified. This is consistent with the expectation that ${\rm Fr}$ increases with $r$ when $l_{\rm drive}$ is set to be constant (Fig.~\ref{fig:FrProfile}).

In previous works \citep[e.g.,][]{kim03, dennis05},
$l_{\rm drive}$ is set to be proportional to the distance from the ICM center: $l_{\rm drive}\approx\alpha r$, where $\alpha$ is a constant coefficient less than unity. Then ${\rm Fr}$ is a decreasing function of $r$, as demonstrated by the dashed lines in Fig.~\ref{fig:FrProfile}. 

There are two possible justifications for $l_{\rm drive}\approx\alpha r$. One is to argue that in the turbulent diffusion coefficient for $D_{\rm edddy} \sim u_{\rm rms} l_{\rm coh}/3$, the velocity coherence length should be a fraction of the pressure scale height $H_{\rm P}$, as in the mixing length theory of convection; the pressure scale height in turn scales with the radius $H_{\rm P} \propto r$ \citep{kim03}. 
In convection, the pressure (or more correctly, the entropy) scale height sets a natural length-scale, since that sets the length scale over which buoyant forces act. However, for extrinsically driven turbulence, the direct scale of driving is what matters, and in our simulations we have set $l$ to be constant. A fluid element at radius $r$ can mix with fluid elements at all radii within an eddy size $l$, not just with fluid elements within the entropy scale height $\propto r$.

Another argument for $l\approx\alpha r$ might be that local turbulent driving scales with radius. Naively, assuming $l\approx\alpha r$ might appear suitable for turbulence driven by AGN feedback, since the size of buoyantly raising bubbles inflated by AGN jet roughly scales with $r$. However, note that AGN turbulence initiated at large radii can still affect gas at smaller radii. The source of turbulence does not have to be local. Furthermore, AGN are not the only source of turbulence. For ICM turbulence driven by large-scale structure motion including infall of sub-clusters, member galaxies, the characteristic scale of turbulence is much larger and does not scale with $r$. The details of this issue -- which require more observational input -- is beyond the scope of this paper. 

\section{Conclusions}
\label{sec:conclude} 

In this work, we study how the gravitational stratification alters the heating rates due to both turbulent dissipation and turbulent diffusion, and how this affects the contribution of turbulence to thermodynamic energy balance in galaxy clusters. The degree of stratification is quantified by the Froude number, ${\rm Fr} = u_{\rm rms}/(l_{\rm drive} N)$, where $u_{\rm rms}$ is the rms turbulent velocity, $l_{\rm drive}$ is the driving scale, and $N$ is the Brunt-Väisälä frequency. We first model these effects analytically, including a new derivation of the impact of stratification on turbulent dissipation. We then perform numerical simulations where the ICM is stirred by turbulence driven by the spectral forcing scheme, and confront analytic theory with our numerical results. 
Our major conclusions are:

\begin{enumerate}
    \item The efficiency of turbulent dissipation is lowered by gravitational stratification, for ${\rm Fr} \lsim 0.1$. The influence of stratification on the turbulent dissipation is physically explained under the theoretical framework of wave turbulence (section~\ref{sec:stratTurb}), where nonlinear interactions of internal gravity waves lead to a turbulent cascade. The turbulent cascade time increases with stratification, becoming larger than the eddy turnover time. The wave turbulence theory predicts that the ratio between actual dissipation rate $\epsilon$ and the dissipation rate of Kolmogorov turbulence, $\epsilon\sim u_{\rm rms}^3/l_{\rm drive}$, scales linearly with the Froude number (Eq.~\ref{eq:dissScaling}). Our simulations show consistency with this scaling relation (Fig.~\ref{fig:dissMain}; Eq.~\ref{eq:betaFr}) in the strong stratification regime (${\rm Fr}\lesssim0.12$); and for ${\rm Fr}\gtrsim 0.12$, the dissipation rate saturates and returns to the Kolmogorov scaling, $\epsilon\approx u_{\rm rms}^3/l_{\rm drive}$.
    Equivalently, in the strong stratification regime, for a given energy injection rate $\epsilon$, turbulent velocities are {\it larger} $u_{\rm rms} \propto {\rm Fr}^{-1/3}$ (Eq. \ref{eq:vFr}, Fig \ref{fig:dissMain}), due to longer cascade times. Thus, observations which measure $u_{\rm rms}$ but do not take this into account and assume Kolmogorov cascade rates, will erroneously infer turbulent heating rates which are too high, by a factor ${\rm Fr}^{-1}$. Finally, in the strongly stratified regime, since $\epsilon  \sim u_{\rm rms}^{4} /(N l_{\rm drive}^{2})$ (Eq. \ref{eq:dissScaling}), the heating rate is even more sensitive to $u_{\rm rms}$ and $l_{\rm drive}$ compared to the Kolmogorov rate $\epsilon \sim u_{\rm rms}^{3}/l_{\rm drive}$. This significantly increases measurement uncertainties in $\epsilon$, particularly since $l_{\rm drive}$ is poorly constrained.

    \item Gravitational stratification suppresses turbulent diffusion, for ${\rm Fr} \lsim 1$. Turbulent diffusion can be described by mixing length theory, where the diffusion coefficient $D_{\rm eddy}\sim u_{\rm rms}l_{\rm drive}$. In the strongly stratified regime, buoyant oscillations dominate over turbulence in the radial direction. Consequently, the heating due to turbulent diffusion is reduced, and the radial diffusion coefficient $D_{\parallel} \propto {\rm Fr}^{2}$. An analytical model describing the diffusion coefficient (Eq.~\ref{eq:stratDiff}) is obtained by interpolating between the unstratified and stratified regimes. We find the  analytical model can accurately describe the turbulent diffusion rate measured in our simulations (Fig.~\ref{fig:deddyFr}). Our best fit model requires the parameter $c_1$ in Eq.~(\ref{eq:stratDiff}) to be a factor of $\sim16$ larger than the original value calibrated in \citet{weinstock81} for conditions appropriate for the Earth's atmosphere. Thus, suppression of turbulent heat diffusion in the cluster context is considerably more important than previously thought, by an order of magnitude. 
    
    \item  
    The turbulent velocity in the simulated halos is found to be more tangentially biased in the inner regions (Fig.~\ref{fig:ansBetaAll}) suggesting stronger stratification (smaller ${\rm Fr}$) inwards. This is consistent with our assumption that the turbulent driving scale is a constant through the simulation domain.
    Previous works which invoke convective mixing length theory often assume $l_{\rm drive}\propto r$, which would otherwise lead to weaker stratification in the center.

    We suggest
    that the driving scale of such turbulence does not scale with $r$; and gravitational stratification in the cluster context is more important than previously thought: ${\rm Fr}$ can be much smaller in the inner core region. Nonetheless, it should be noted that while the \bv\, frequency $N$ can be determined from observed entropy profiles, and $u_{\rm rms}$ can also be constrained observationally, the driving scale $l_{\rm drive}$ is more difficult to pin down and is the principle uncertainty in determining the Fr profile of clusters. 
    Also note that if thermal conduction is efficient -- although this is unclear --then the \bv\, frequency $N$ is proportional to the logarithmic temperature gradient, rather than the logarithmic entropy gradient, which typically lowers it by a factor of $\sim 2$ \citep{sharma09-BV,ruszkowski10}. 

Overall, our results suggest that suppression of turbulent diffusion by stratification is very strong, and important over a large radial range in cluster cores (the radial turbulent diffusion coefficient $D_{\parallel} \propto {\rm Fr}^{2}$, for ${\rm Fr} \lsim 1$). It cannot be ignored. It should also be noted that the turbulent diffusion of other passive scalars (e.g., metallicity) will be similarly affected. Suppression of turbulent dissipation follows a linear scaling ($\epsilon \propto {\rm Fr}$, for ${\rm Fr} \lsim 0.1$). 
Typically, only the innermost parts of ICM (e.g., in the inner $\sim10$ kpc for the fiducial model in Fig.~\ref{fig:FrProfile}) would be affected. Therefore,  the results of \citet{2014Natur.515...85Z} are not likely to be significantly affected by gravitational stratification.
Nonetheless -- modulo assumptions about driving scales -- suppression can be strong in the innermost regions, and it is important to keep in mind. These effects may also be important in other contexts, e.g., the circumgalactic medium, and stellar and planetary atmospheres. 
\end{enumerate}

\section*{Acknowledgements}

We thank Rajsekhar Mohapatra, Annick Pouquet, Mark Voit, and Max Gronke for helpful conversations. This research was supported in part by the National Science Foundation under Grant No. NSF PHY- 1748958 to KITP; we thank the organizers and participants of the KITP ``Fundamentals of Gaseous Halos" workshop for helpful discussions.  CW and SPO acknowledge NASA grant 19-ATP19-0205 and NSF grant AST-1911198 for support. MR acknowledges NSF grant AST 2009227, NASA grant 80NSSC20K1583, and Forschungsstipendium from the Max-Planck-Institut für Astrophysik in Garching, Germany.

\section*{Data Availability}
The data underlying this article will be shared on reasonable request to the corresponding author.



\bibliographystyle{mnras}
\bibliography{main} 








\bsp	
\label{lastpage}
\end{document}